\documentclass[structabstract]{aa}

\usepackage{natbib}
\usepackage{graphicx}
\usepackage{epstopdf}
\usepackage{color}
\usepackage{hyperref}
\hypersetup{colorlinks, citecolor=blue, filecolor=black, linkcolor=black, urlcolor=black}

\newcommand{\RNum}[1]{\uppercase\expandafter{\romannumeral #1\relax}}

\usepackage{subfigure}

\begin{document}

\title{Improved methods for determining the kinematics of coronal mass ejections and coronal waves}

\titlerunning{Improved methods for determining the kinematics of CMEs and coronal waves}
\authorrunning{Byrne et al.}

\author{J.\,P.\,Byrne\inst{1}
	\and D.\,M.\,Long\inst{2}
	\and P.\,T.\,Gallagher\inst{3}
	\and D.\,S.\,Bloomfield\inst{3}
	\and S.\,A.\,Maloney\inst{4}
	\and R.\,T.\,J.\,McAteer\inst{5}
	\and H.\,Morgan\inst{6,1,7}
	\and S.\,R.\,Habbal\inst{1}
	}
\institute{Institute for Astronomy, University of Hawai'i, 2680 Woodlawn Drive, Honolulu, HI 96822, USA.\\
		\email{jbyrne@ifa.hawaii.edu}
		\and
		UCL--Mullard Space  Science Laboratory, Holmbury St. Mary, Dorking, Surrey, RH5 6NT, UK.
		\and 
		School of Physics, Trinity College Dublin, College Green, Dublin 2, Ireland.
		\and
		Skytek, 51/52 Fitzwilliam Square West, Dublin 2, Ireland
		\and
		Department of Astronomy, New Mexico State University, Las Cruces, NM 88003-8001, USA.
		\and
		Sefydliad Mathemateg a Ffiseg, Prifysgol Aberystwyth, Ceredigion, Cymru, SY23 3BZ, UK.
		\and
		Coleg Cymraeg Cenedlaethol, Y Llwyfan, Ffordd y Coleg, Caerfyrddin, Cymru, SA31 3EQ, UK.
		}

\date{Received 1 February 2013 / Accepted 26 July 2013}
\abstract
{The study of solar eruptive events and associated phenomena is of great importance in the context of solar and heliophysics. Coronal mass ejections (CMEs) and coronal waves are energetic manifestations of the restructuring of the solar magnetic field and mass motion of the plasma. Characterising this motion is vital for deriving the dynamics of these events and thus understanding the physics driving their initiation and propagation. The development and use of appropriate methods for measuring event kinematics is therefore imperative.} 
{Traditional approaches to the study of CME and coronal wave kinematics do not return wholly accurate nor robust estimates of the true event kinematics and associated uncertainties. We highlight the drawbacks of these approaches, and demonstrate improved methods for accurate and reliable determination of the kinematics.}
{The Savitzky-Golay filter is demonstrated as a more appropriate fitting technique for CME and coronal wave studies, and a residual resampling bootstrap technique is demonstrated as a statistically rigorous method for the determination of kinematic error estimates and goodness-of-fit tests.}
{It is shown that the scatter on distance-time measurements of small sample size can significantly limit the ability to derive accurate and reliable kinematics. This may be overcome by (\emph{i}) increasing measurement precision and sampling cadence, and (\emph{ii}) applying robust methods for deriving the kinematics and reliably determining their associated uncertainties. If a priori knowledge exists and a pre-determined model form for the kinematics is available (or indeed any justified fitting-form to be tested against the data), then its precision can be examined using a bootstrapping technique to determine the confidence interval associated with the model/fitting parameters.}
{Improved methods for determining the kinematics of CMEs and coronal waves are demonstrated to great effect, overcoming many issues highlighted in traditional numerical differencing and error propagation techniques.}



\maketitle

%

\section{Introduction}
\label{sect_intro}

Coronal mass ejections (CMEs) and coronal waves (commonly known as ``EIT waves") are large-scale manifestations of solar activity that indicate a restructuring of the global solar magnetic field. These phenomena involve the mass motion of plasma through the solar corona, with energies on the order of $10^{25}$\,J for CMEs \citep{2004JGRA..10910104E}, and upwards of $10^{18}$\,J for coronal waves \citep{2005ApJ...633L.145B}. CME speeds range from about 20 to $>$$2500$\,km\,s$^{-1}$ \citep{2004JGRA..10907105Y}, most typically moving at speeds similar to those of coronal waves, which range from 50 to $>$$700$\,km\,s$^{-1}$ \citep{2009ApJS..183..225T}. Observational catalogues of these events have been compiled from over $\sim$20\,years of observations, with the aim of characterising their physical properties in order to better understand the dynamics of their initiation and propagation \citep[see some recent reviews by][]{2011ASSL..376.....H, 2011SSRv..158..365G, 2012SoPh..tmp...93P, 2012LRSP....9....3W}, and form an integral part of such efforts as the Solar Dynamics Observatory Feature Finding Team \citep[\emph{SDO} FFT;][]{2012SoPh..275...79M}, the Heliophysics Event Knowledgebase \citep[HEK;][]{2012SoPh..275...67H}, and the HELiophysics Integrated Observatory \citep[HELIO;][]{2011AdSpR..47.2235B} . CME dynamics, in particular, are of great interest in a space weather context \citep[e.g.,][]{2005A&A...440..373H, 2010heliophysics, SWE:SWE493}. Therefore, methods for determining their kinematics with improved accuracy are extremely important if scientists are to become skilled at predicting their behaviour and arrival times \citep[see, for example, efforts by][]{2004Natur.432...78P, 2005AnGeo..23.1033S, 2006ApJ...652.1747C, 2010NatCo...1E..74B}. \citet{2012ApJ...749...57T} also highlight the importance of characterising event kinematics as accurately as possible, in order to understand the effects of drag and CME-CME interactions and the subsequent implications for the underlying magnetic field structures involved.

In order to observe and characterise the motion of the bulk plasma of these events, a method of image-differencing is traditionally used, whereby a preceding image is subtracted from a leading image to highlight moving features. However, this approach enhances relative rather than actual motion and is prone to spatiotemporal crosstalk and user-dependent bias. More recent work has used single-image processing techniques such as multiscale filters \citep{2008SoPh..248..457Y, 2009A&A...495..325B, 2011AdSpR..47.2118G} and robust automated approaches \citep[e.g.,][]{2011A&A...531A..42L, 2012SoPh..276..479P, 2012ApJ...752..145B} to overcome these issues and reveal the true physical characteristics of the events. Thus, by accurately tracking the position of a feature over time, it is possible to determine the kinematics of the event. 

The true physical nature of coronal waves is not fully understood, with two main competing theories: that they are indeed waves \citep[e.g.,][]{2010ApJ...716L..57V, 2012ApJ...754....7S}, or that they are signatures of magnetic field restructuring during a CME eruption \citep[e.g.,][]{2011ApJ...738..167S,2011ApJ...732L..20C}. Their kinematic behaviour has been proposed as one of the main discriminators between these competing theories, with the relatively high velocities measured thus far suggesting a wave interpretation to be appropriate \citep{2011A&A...532A.151W, 2012ApJ...753..112Z}. Similarly, the low-coronal kinematics of CMEs may be used to discriminate between eruption mechanisms \citep[see, for example,][and the CME models discussed therein]{2010A&A...516A..44L}. However, \citet{2007ApJ...657.1117W} demonstrate that the errors in CME acceleration values can be of the same order as the accelerations typically measured, making this task difficult. This has led to many statistical studies employing large numbers of events to try and determine a general form for typical CME motion \citep[e.g.,][]{2000GeoRL..27..145G, 2003AdSpR..32.2637D, 2006ApJ...649.1100Z}. However, individual events need to be studied with rigour in order to satisfactorily derive the kinematics and gain insight to the physics at play.

A variety of different mathematical techniques exist for deriving the kinematics of transient features, most being based upon some form of numerical differentiation of the distance-time measurements and/or the fitting of a pre-assumed model function. While such techniques may appear mathematically sound, some of them are not necessarily applicable to the derivation of kinematics for these specific forms of events and can produce spurious results. \citet{2010ApJ...712.1410T} demonstrated the applicability of an inversion technique (developed by \citealt{2005SoPh..227..299K}) to overcome some of these issues. They highlight its effectiveness at reducing the errors on three CME/flare studies compared to a standard spline-fit approach, and so the technique shows promise if it can be applied robustly across a variety of events.

Here, we further explore improved methods for overcoming the issues outlined above to robustly determine the kinematics of CMEs and coronal waves with improved accuracy and reliability. Simulations of the drawbacks of a standard numerical derivative are presented in Sect.~\ref{sect:simul1}. In Sect.~\ref{sect:bootstrapping}, we outline a more appropriate method for inspecting the  kinematics of CMEs and coronal waves, as applied to model data. In Sect.~\ref{sect:case_studies}, some real-data cases are inspected via these methods, motivated by the proposed treatment of data from the new coronal image processing CME catalogue \citep[CORIMP;][]{2012ApJ...752..144M, 2012ApJ...752..145B} and coronal pulse identification and tracking algorithm catalogue \citep[CorPITA;][]{2011A&A...531A..42L}. The main conclusions and future work are discussed in Sect.~\ref{sect:conclusions}.

\section{Simulated data}
\label{sect:simul1}

\begin{figure*}[!t]
\centering
\subfigure{\includegraphics[scale=0.5, trim=0 90 0 0]{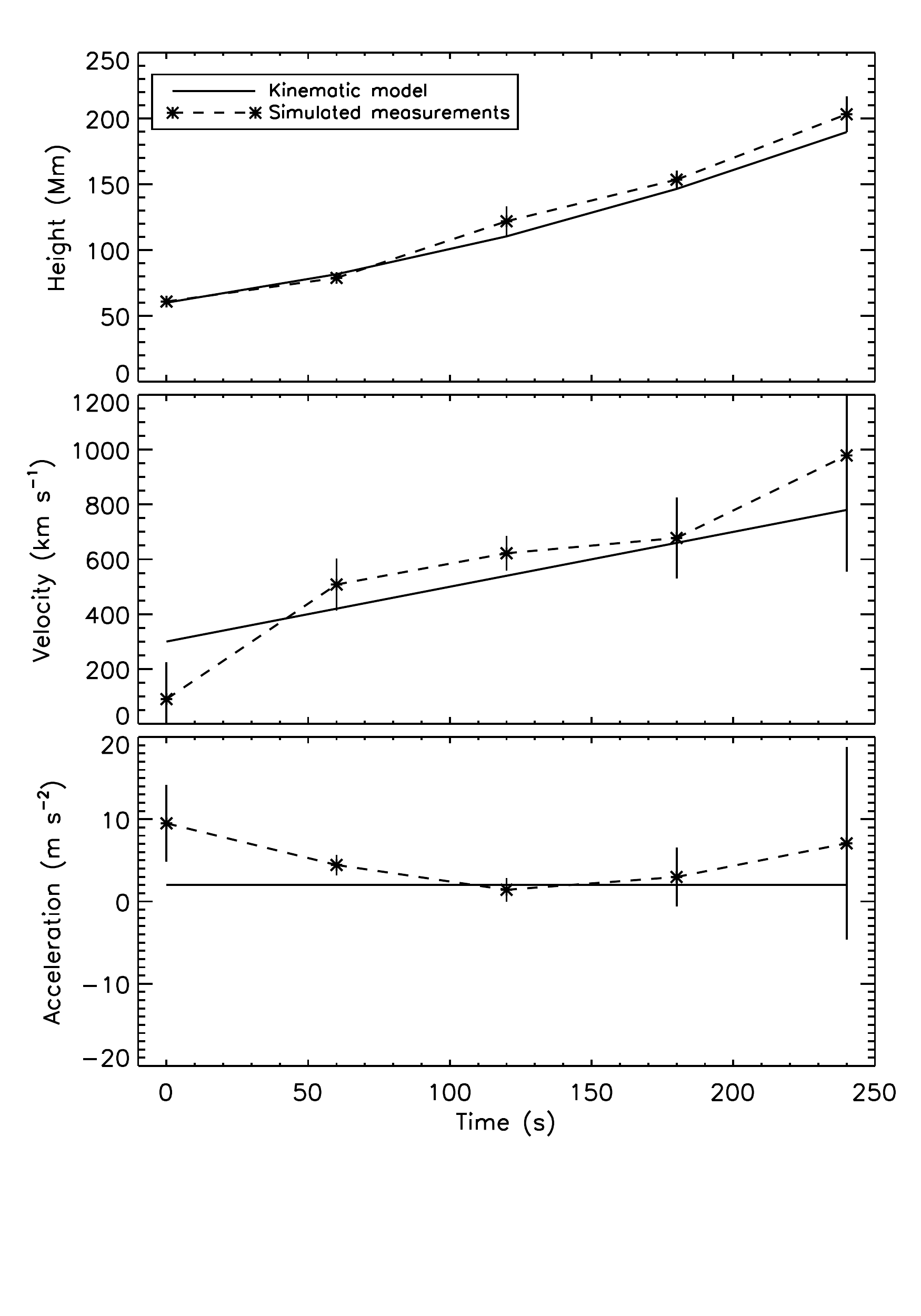}}
\subfigure{\includegraphics[scale=0.5, trim=0 90 0 0]{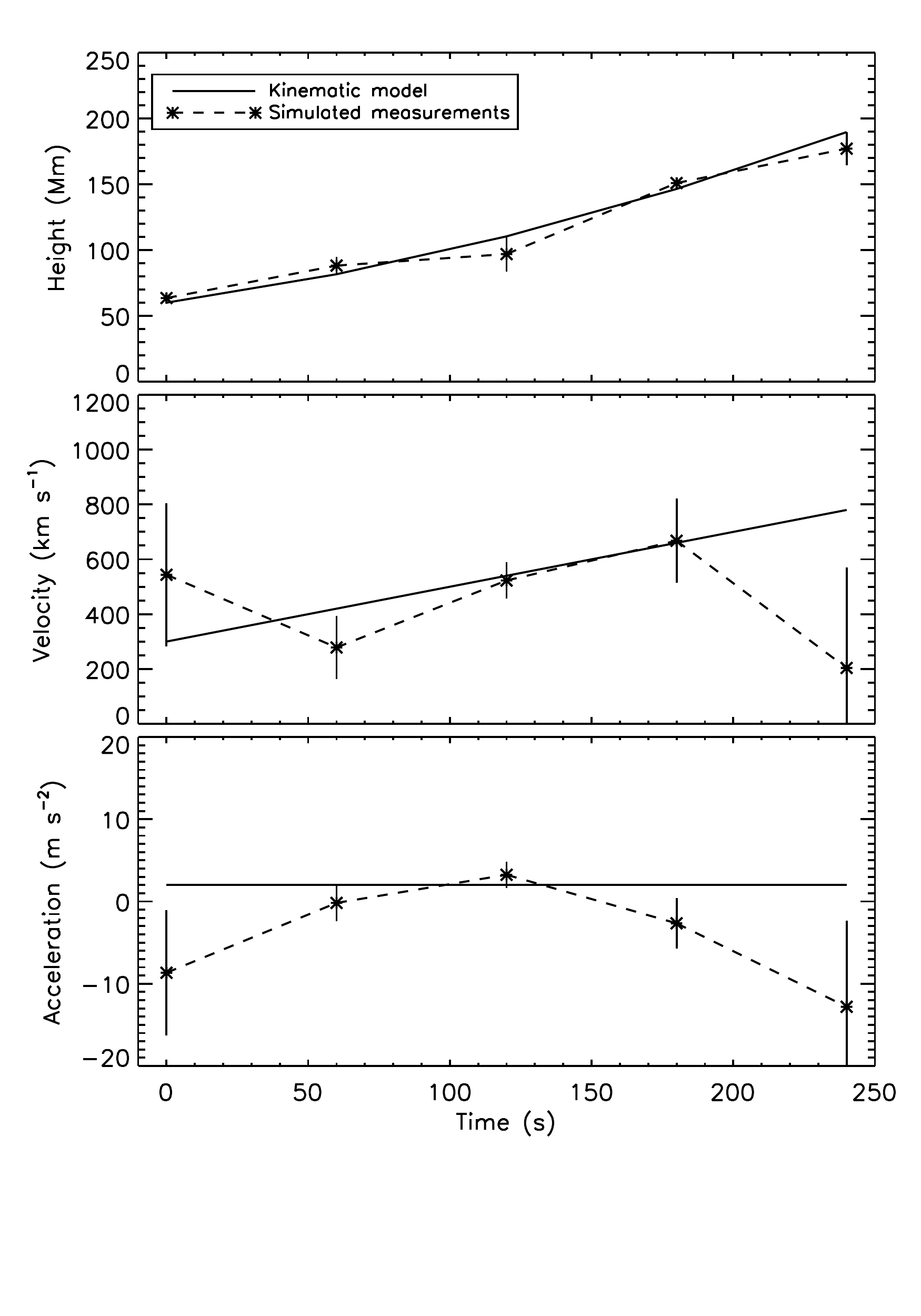}}
\caption{Kinematic model for a CME with constant acceleration $2$\,m\,s$^{-2}$ and initial velocity $300$\,km\,s$^{-1}$, and two data simulations of how the resulting profiles for different scatters on the height-time measurements (\emph{top panels}) behave when 3-point Lagrangian interpolation is used to derive the velocities (\emph{middle panels}) and accelerations (\emph{bottom panels}). Both simulated cases were produced by adding scatter to the height measurements, via a normally-distributed random number generator with a standard deviation of 10\% of the model height at each time-step. The different instances of scatter shown here produce completely opposing trends in the accelerations, with the errorbars failing to appropriately overlap the model, therefore belying the true trend.}
\label{sim_vels_thesis}
\end{figure*}

When presented with relatively low sampling of the data, it is generally found that the simplest differentiation techniques are not applicable. The forward and reverse differencing techniques act to shift the kinematic profiles by one time step, which is substantial enough to be of concern here (i.e., they derive a result at the current time step, based on the change from the preceding or proceeding time step). Centre differencing employs the two neighbouring data points of the point under examination, and so is a better indication of the result at that time step, but it fails at the endpoints. In any case, these techniques should not be employed when the spacing of the data points is unequal (i.e., when the cadence, $\delta t$, is not constant). Therefore, the 3-point Lagrangian interpolation technique is often used (as in \textsc{deriv.pro} in IDL), which includes the endpoints and has an associated error propagation formulation, which for the velocity is given by,
\begin{equation}
\sigma_{v_1}^2 \,=\, \frac{\sigma_{r(t_2)}^2+\sigma_{r(t_0)}^2}{(t_2-t_0)^2} + v^2 \frac{\sigma_{t_2}^2+\sigma_{t_0}^2}{(t_2-t_0)^2}\ ,
\label{vel_err}
\end{equation}
and similarly for acceleration (as in \textsc{derivsig.pro} in IDL, from \citealt{2003drea.book.....B}). The endpoint errors are derived from a weighting of the preceding or proceeding two data points and are therefore larger, reflecting the unknown nature of the trend beyond these points.
Although the 3-point Lagrangian is mathematically sound, its application to CME and coronal wave kinematics proves problematic. The main drawbacks, which will be discussed in detail in the following subsections, are two-fold:
\begin{enumerate}
\item The scatter in the measurements, especially across low-cadence sampling, can cause the numerical derivatives to become untrustworthy and even misleading compared to the actual trends of the kinematic data.
\item The error-propagation formulation results in a misleading uncertainty on the velocity and acceleration profiles, whereby the errorbars counter-intuitively increase in size when the cadence itself is increased (i.e., when more frequent observations are made).
\end{enumerate}
In our treatment of both phenomena here, we use cases of simulated CME and coronal wave measurements interchangeably to demonstrate both constant and non-constant acceleration profiles with varying measurement scatter and cadence.

\begin{figure}[!t]
\begin{center}
\includegraphics[width = 0.45\textwidth]{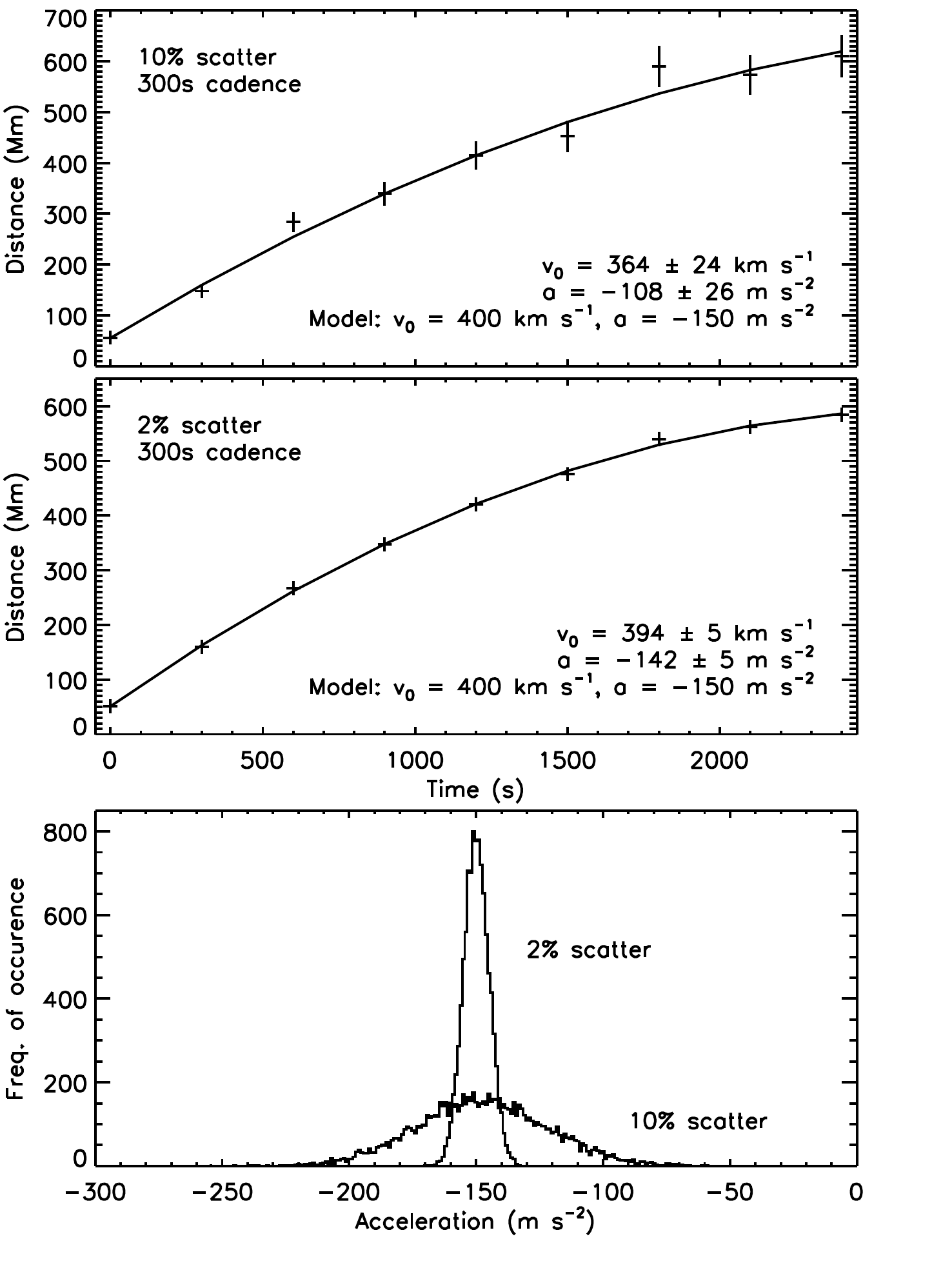}
\caption{Simulated distance-time measurements with 3$\sigma$ scatters of $\pm$\,10\% (\emph{top}), and $\pm$\,2\% (\emph{middle}) of the model value, at a fixed cadence of $300$\,s, and the resulting quadratic fit and $v_0$ and $a$ parameters. The reduced scatter increases the precision for obtaining the true kinematics, as demonstrated by the different distributions of derived acceleration fit parameters from 10\,000 runs of the simulation (\emph{bottom}).}
\label{noise_hist_weight}
\end{center}
\end{figure}

\begin{figure}[!t]
\begin{center}
\includegraphics[width = 0.45\textwidth]{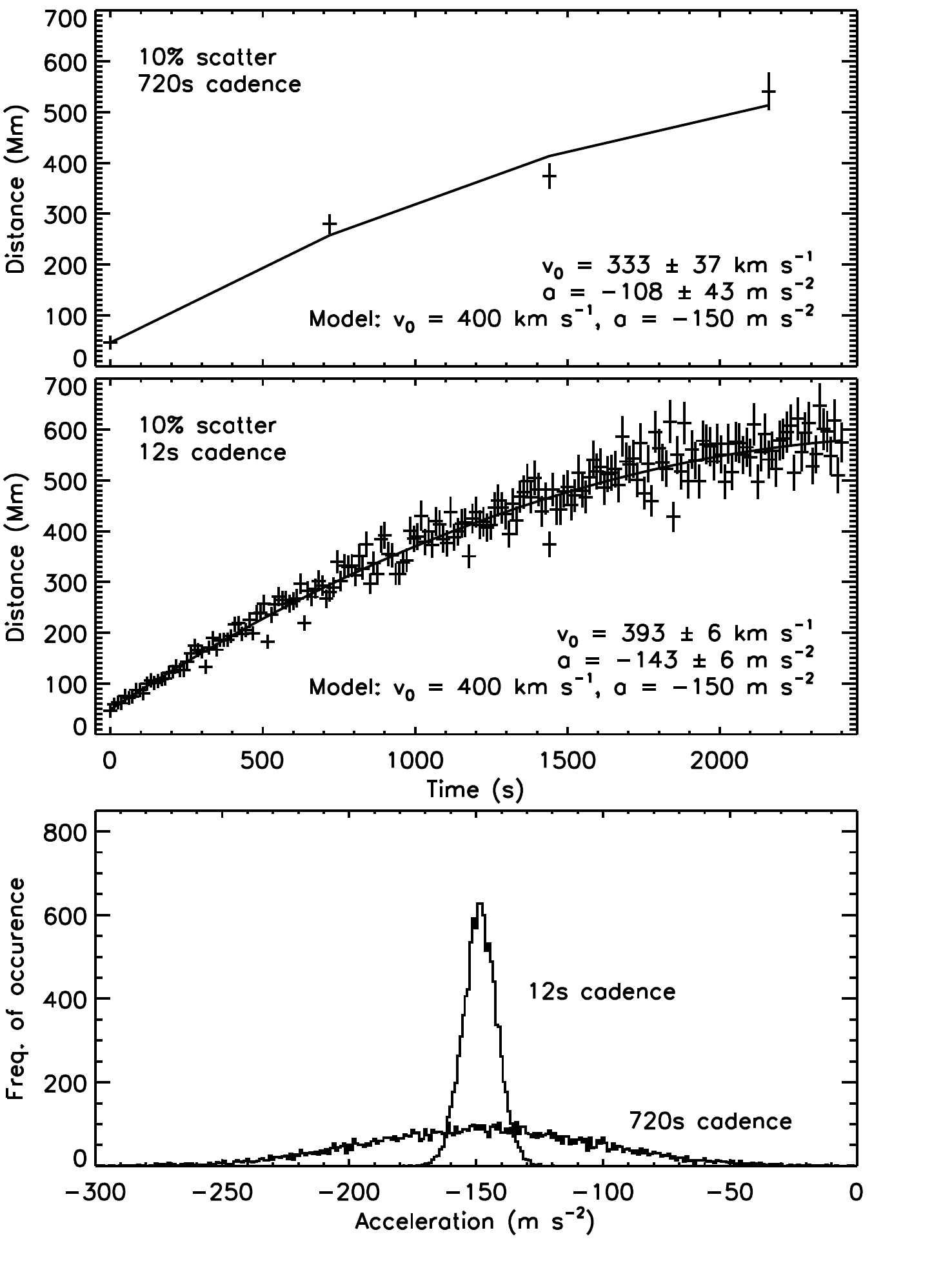}
\caption{Simulated distance-time measurements for sampling cadences of 720\,s (EIT; \emph{top}) and 12\,s (AIA; \emph{middle}), at a fixed scatter of $\pm\,10\%$, and the resulting quadratic fit and $v_0$ and $a$ parameters. The increased cadence offers better precision for obtaining the true kinematics, as demonstrated by the different distributions of derived acceleration fit parameters from 10\,000 runs of the simulation (\emph{bottom}).}
\label{cad_hist_weight}
\end{center}
\end{figure}

\begin{figure}[!b]
\begin{center}
\includegraphics[scale=0.53, trim=20 10 0 20]{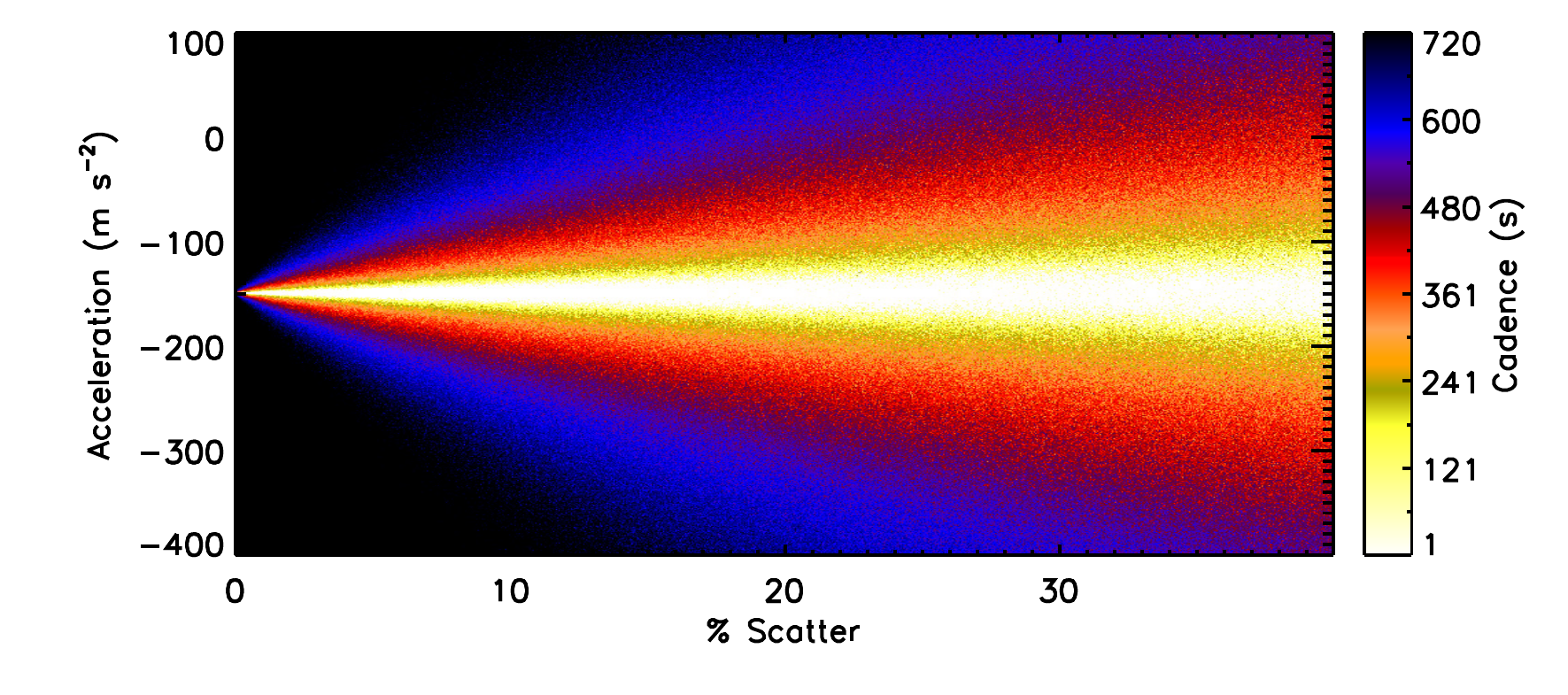}
\caption{Simulation of the derived accelerations from the model fits to Equation~\ref{eqn:const_a} (with examples shown in Figs.~\ref{noise_hist_weight} and \ref{cad_hist_weight}), for scatters of 0\,--\,40\% and cadences of 1\,--\,720\,s. As shown, a decrease in both the data scatter and the cadence time improves the chances of obtaining the correct acceleration value, being $-150$\,m\,s$^{-2}$ in this model coronal wave case.}
\label{noise_test_image}
\end{center}
\end{figure}

\subsection{Effect of measurement scatter on deriving kinematics}
\label{subsect:noise}

As an example of the effect of measurement scatter in the data, we first simulate a simple height-time profile of a CME that propagates according to the simple quadratic equation,
\begin{equation}
\label{eqn:const_a}
r(t) = r_0 + v_0 t + \frac{1}{2}a t^2 \ ,
\end{equation}
where $r_0=60$\,Mm is the initial height, $v_0=300$\,km\,s$^{-1}$ is the initial velocity, and $a=2$\,m\,s$^{-2}$ is the acceleration of the CME. Measurement scatter is applied to the height-time points via a normally-distributed random number generator with a standard deviation of 10\% of the model height at each time step. This is to simply convey the fact that the events that will be studied are transient phenomena that become increasingly difficult to discern in the images as their signal-to-noise ratios decrease and their structure often becomes disjoint. Each measurement is assigned an errorbar sufficient to overlap the true height-time profile. Even in this simple case of constant acceleration, various instances of randomised measurement scatter result in erroneous trends in the velocity and acceleration profiles, despite using the proper error treatment prescribed by the 3-point Lagrangian interpolation technique (Equation~\ref{vel_err}). The endpoint errors are derived from a weighting of the preceding or proceeding two data points and are therefore larger, reflecting the unknown nature of the trend beyond these points.

Two examples are shown in the left and right panels of Fig.~\ref{sim_vels_thesis}, where completely opposing acceleration trends are determined for different samplings of the same height-time data set. This indicates that the nature of the scatter in the samples is not satisfactorily reflected in the derived kinematics and their associated errorbars. At the very least, the kinematic uncertainties should be expected to overlap the truth in each plot so that it remains a valid solution. A possible workaround is that instead of trusting the endpoints they simply be removed. Figure~\ref{sim_vels_thesis} would then show three data points for velocity and one data point for acceleration, reducing the biased trends. However, when dealing with low-number samples it would be better not to have to remove data points.

The effects of scatter on the distance-time measurements of a coronal wave were also examined, demonstrated here for the case of a constant acceleration event. The wave motion is modelled by Equation~\ref{eqn:const_a}, where $r_0\,=\,50$\,Mm is the initial distance of the wave from the source, $v_0\,=\,400$\,km\,s$^{-1}$ is the initial velocity of the wave, and $a\,=\,-150$\,m\,s$^{-2}$ is the acceleration of the wave. Figure~\ref{noise_hist_weight} shows the derived kinematics for the simulated distance-time measurements with random scatter added, shown here for 3$\sigma$ limits of 10\% (top panel) and 2\% (middle panel). Errorbars are applied with magnitude equal to the measurement scatter percentage. A second-order polynomial (quadratic) is then fit to each data set to test how the scatter affects the precision of the derived kinematics, even in this idealised case of knowing the underlying form of the data. The increased scatter acts to smooth out the true kinematics, as demonstrated by the different distributions of derived accelerations from 10\,000 runs of the simulation (bottom panel).

\subsection{Effect of sampling cadence on deriving kinematics}
\label{subsect:cadence}

As an example of the effect of cadence, we first simulate again the constant-acceleration profile of a coronal wave (as in Equation~\ref{eqn:const_a} and Fig.~\ref{noise_hist_weight}). The simulated distances were sampled at cadences similar to the Solar \& Heliospheric Observatory \citep[\emph{SOHO};][]{1995SoPh..162....1D} Extreme-ultraviolet Imaging Telescope (EIT; 12\,min) and the Solar Dynamics Observatory \citep[\emph{SDO};][]{2012SoPh..275....3P} Atmospheric Imaging Assembly (AIA; 12\,s), with a $3\sigma$ scatter of 10\% of the model distance. Errorbars are again applied with magnitude equal to the measurement scatter percentage. Figure~\ref{cad_hist_weight} shows examples of these, with a quadratic fit to the sample measurements to test the effect on the derived kinematics. It is clear that the higher-cadence data best resolves the true kinematic profile, providing an accurate estimation of the wave velocity and acceleration. These results are consistent with the observations made by both \citet{2008ApJ...680L..81L} and \citet{2009ApJ...707..503M}, and show that the effects of image cadence must be accounted for when determining the kinematics of a coronal wave.

\begin{figure*}[!ht]
\centering
\subfigure{\includegraphics[scale=0.5, trim=0 55 0 50]{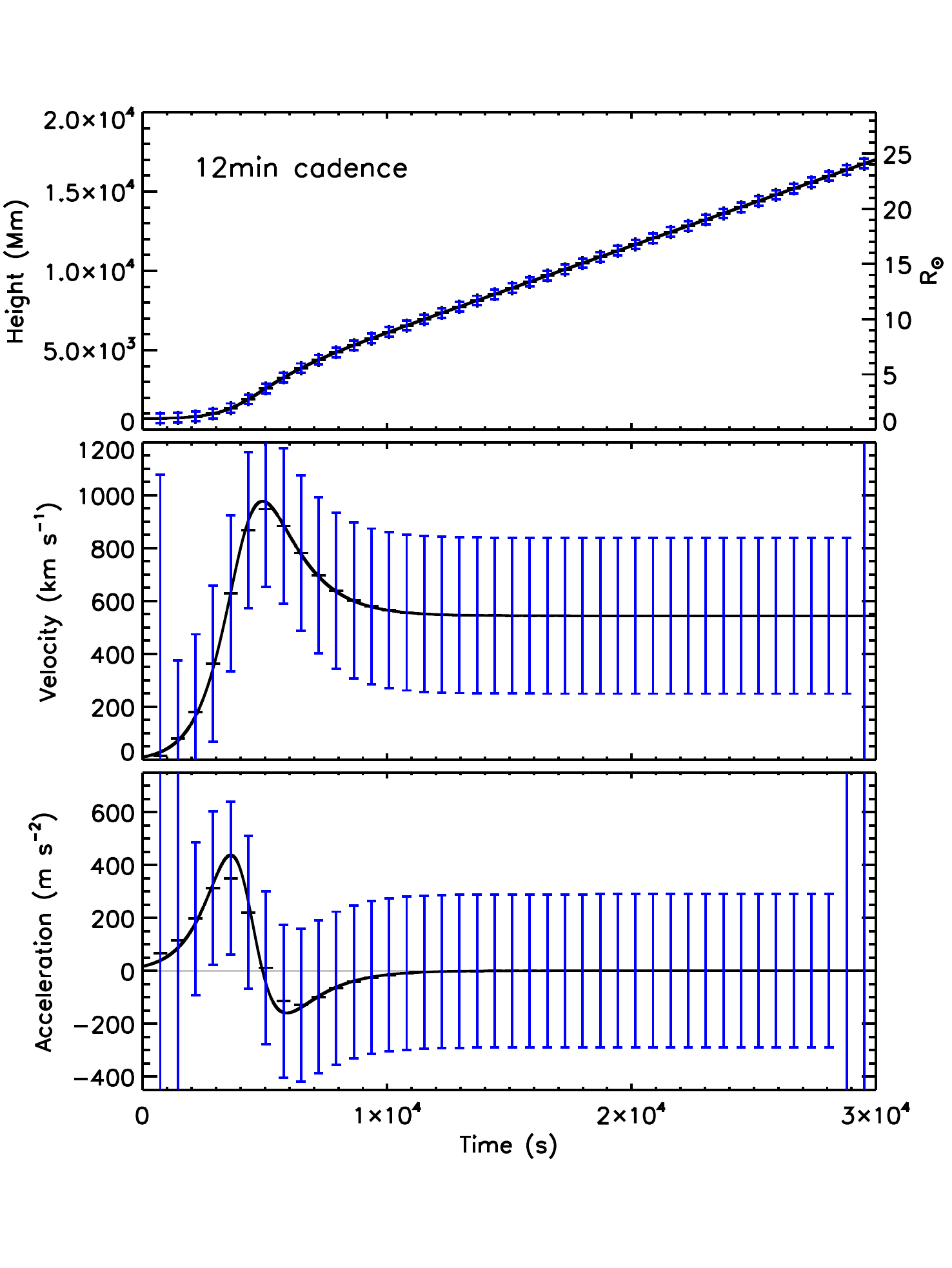}}
\subfigure{\includegraphics[scale=0.5, trim=0 55 0 50]{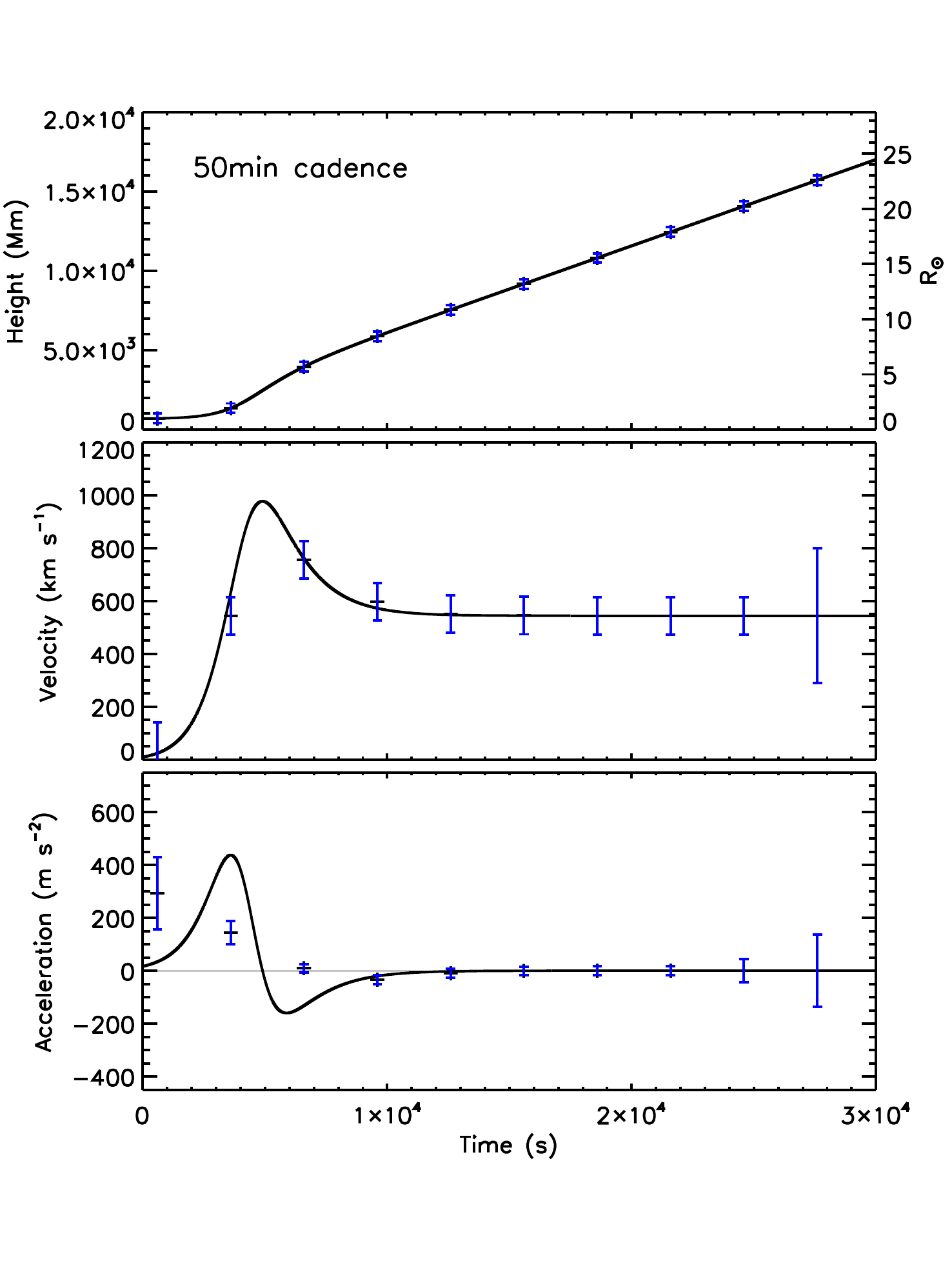}}
\subfigure{\includegraphics[scale=0.5, trim=0 20 0 25]{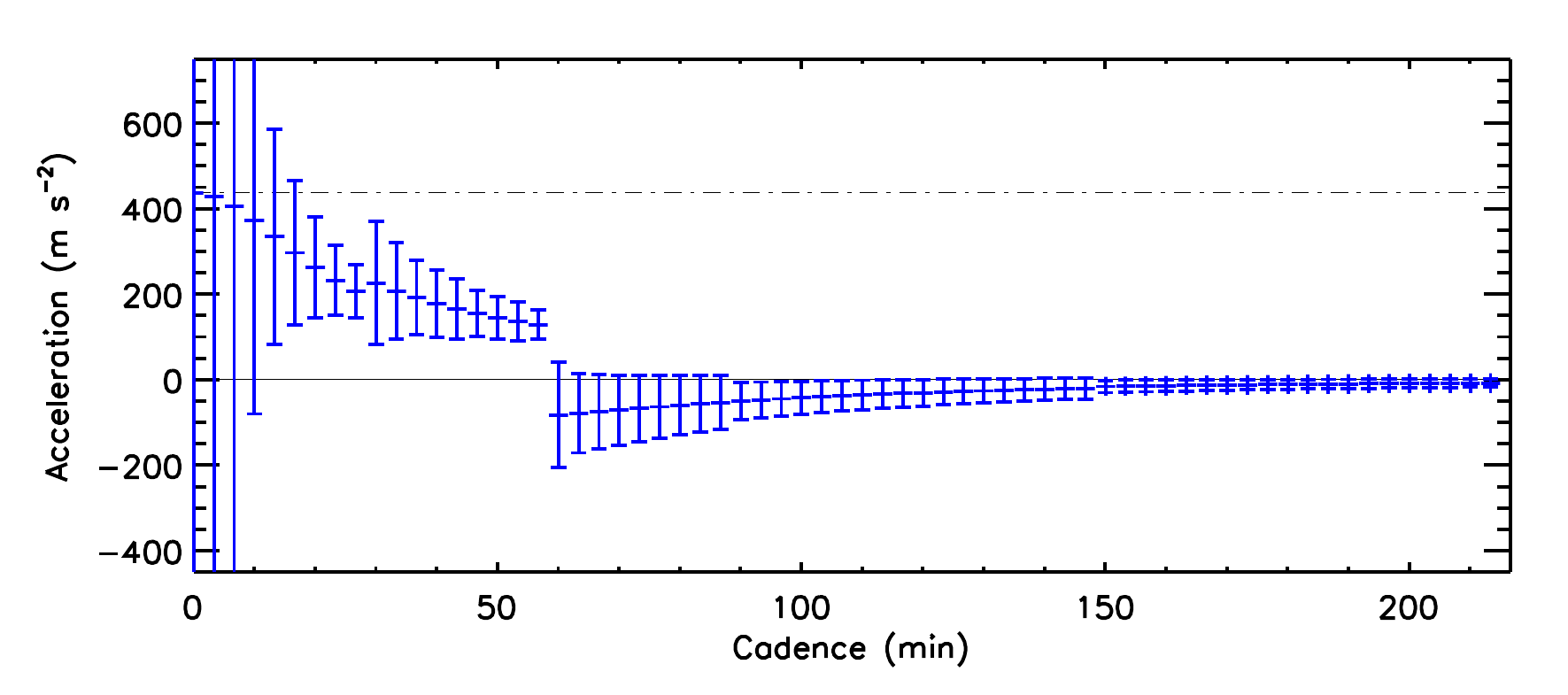}}
\caption{Demonstration of the effects of cadence on the error propagation according to the 3-point Lagrangian interpolation. A kinematic model for a CME with non-constant acceleration peaking at 437\,m\,s$^{-2}$ is tested for varying cadences, without any scatter. The top left plots show the height, velocity, and acceleration profiles for data sampled at 12\,minute cadence. The top right plots show the same data sampled at 50\,minute cadence. Note how the errorbars of the higher cadence measurements are counter-intuitively larger than the lower cadence measurements, even though the higher sampling rate better reveals the true kinematic trend. The bottom plot shows the derived peak acceleration against cadence, where the dot-dashed line indicates the true value of 437\,m\,s$^{-2}$. The errorbars are shown to reduce in magnitude (implying greater precision) even though the derived acceleration at lower cadence is less accurate.}
\label{fig_cadence_hva}
\end{figure*}

In Fig.~\ref{noise_test_image} the combined effects of measurement scatter and sampling cadence are simulated for the model coronal wave case. The image shows the result of plotting the distribution of acceleration fit parameters against all variations of scatter from $0$\,--\,$40\%$, at all variations of cadence from $1$\,--$720$\,s. Essentially the plots of the bottom panels of Figs.~\ref{noise_hist_weight} and \ref{cad_hist_weight} represent slices through the corresponding locations of Fig.~\ref{noise_test_image}. This demonstrates that the reduction of measurement scatter and increase of sampling cadence together are required to improve the accuracy of the derived kinematics. However, measurement scatter is less problematic for higher cadence data sets.

We next simulate a non-constant acceleration profile for a typical fast CME via the following equations,
\begin{eqnarray}
h(t)\,=&\,\sqrt{2s}\,t\tan^{-1}\left(\frac{e^{t/2s}}{\sqrt{2s}}\right) \ , \\
v(t)\,=&\,\sqrt{2s}\tan^{-1}\left(\frac{e^{t/2s}}{\sqrt{2s}}\right)+\frac{e^{t/2s}t}{e^{t/s}+2s} \ , \\
a(t)\,=&\,\frac{e^{t/2s}\left(2s\left(t+4s\right)-e^{t/s}\left(t-4s\right)\right)}{2s\left(e^{t/s}+2s\right)^2}\ ,
\label{eqn:nonconst_a}
\end{eqnarray}
where $s$ is a constant scaling factor. The acceleration profile exhibits an initial peak followed by a deceleration and then levels to zero. This is similar to a general impulsive CME that undergoes an initial high-acceleration eruptive phase before it decelerates to match the solar wind speed during its propagation phase. A model CME height-time profile is generated, enabling synthetic observation samples to be taken at different cadences.

We investigate the effect of observational cadence on the derivation of the kinematics and associated errorbars using the standard 3-point Lagrangian interpolation. In the first instance, fixed errorbars of $\pm\,300$\,Mm are applied to the height-time measurements without any scatter. This is useful to simply test the effects of the cadence on the derived velocity and acceleration profiles and their associated errors. The left and right plots of Fig.~\ref{fig_cadence_hva} show the model height, velocity, and acceleration profiles sampled at cadences of 12 and 50\,minutes, respectively. As the cadence is reduced (i.e., the time interval between observations is increased), the errorbars become smaller due to the inverse dependence of the Lagrangian error terms on the time between the data points $\Delta t^{-2}$ (see Equation~\ref{vel_err}). However, reducing the cadence reduces the resolution at which the acceleration peak is detectable, and so the acceleration profile is less well characterised. Conversely, the errorbars become unrealistically large for very high-cadence observations even though the measurements better reveal the true trends of the kinematic profiles, as demonstrated in the bottom plot of Fig.~\ref{fig_cadence_hva}. This fundamentally implies that the errorbars do not truly reflect the uncertainty on the data at a given cadence, and are in fact redundant for these cases.

It is clear that the variation in both scatter and imaging cadence can strongly influence the derived kinematics of a CME or coronal wave, and the commonly used 3-point Lagrangian technique does not return useful estimates of the associated uncertainty. Furthermore, the quantification of uncertainty on the physical measurements themselves is extremely non-trivial. Uncertainties exist that are attributable to the unknown physical mechanisms operating in the phenomena being studied, as well as the uncertainties involved in the techniques used for the analysis, making a robust error estimate practically impossible. This is especially true for automated routines designed to detect faint and transient phenomena such as CMEs and coronal waves. However, it is possible to use other techniques in an effort to overcome these issues and produce a more statistically sound method for dealing with these types of measurements.

\section{Bootstrapping: a resampling method}
\label{sect:bootstrapping}

When trying to determine an estimator for a particular parameter of interest and subsequently evaluate the accuracy of that estimator, a small sample size is immediately limiting. Therefore, in order to approximate the behaviour of the true distribution, techniques based on resampling methods have been developed. These work by resampling the data enough times to generate a maximum likelihood estimator of the distribution. Bootstrapping is one such technique, first introduced by \citet{Efron:1979p1831} and more recently described in, e.g., \citet{1994.book.Efron} and \citet{Chernick1999}. 

\begin{figure}
\begin{center}
\includegraphics[scale=0.45, trim=20 50 0 0, clip=true]{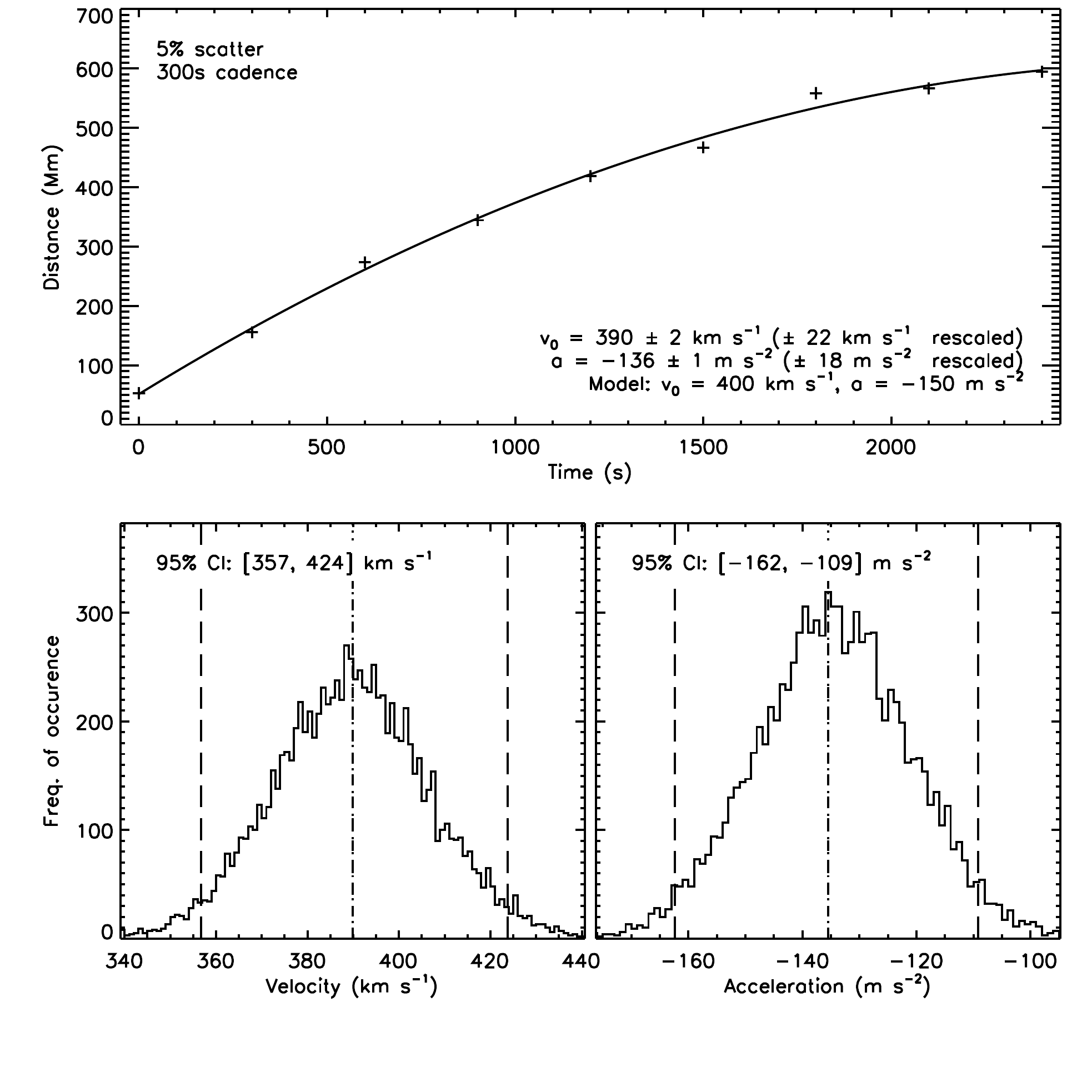}
\caption{Top panel: initial fit of Equation~\ref{eqn:const_a} to simulated coronal wave distance-time measurements, with 5\% scatter and 300\,s cadence. Errorbars of $\pm1.1$\,Mm are assigned to the measurements, based on the 1.5\,arcsec resolution of the imager. The fit parameters are quoted with $1\sigma$ uncertainties, with the rescaled uncertainties shown in brackets. Bottom panels: histograms of the initial velocity and acceleration values derived using the bootstrapping technique. The mean and 95\% confidence interval are indicated by the dot-dashed and dashed lines respectively. Bootstrapping provides a distribution of fitting parameters that is unattainable via a standard single-fit to data when unknown sources of uncertainty exist.}
\label{cad_boot_weight}
\end{center}
\end{figure}

In the cases of CME and coronal wave observations, bootstrapping techniques can prove very useful for determining the uncertainty on the derived kinematic model parameters. The implementation of the residual resampling bootstrapping scheme is as follows:
\begin{enumerate}
\item An initial fit to the data $y$ is obtained, yielding the model fit $\hat{y}$ with parameters $\vec{p}$.
\item The residuals of the fit are calculated as $\epsilon = y - \hat{y}$.
\item The residuals are randomly resampled with replacement to give $\epsilon^*$.
\item The model is then fit to a new data vector $y^* = \hat{y} + \epsilon^*$ and the parameters $\vec{p}^*$ stored.
\item Steps 3--4 are repeated many times (e.g., 10\,000).
\item Confidence intervals on the parameters are determined from the resulting distributions.
\end{enumerate}
This technique was used to fit a quadratic model to the simulated coronal wave moving with constant acceleration (Equation~\ref{eqn:const_a}) and to the simulated CME moving with non-constant acceleration (Equation~\ref{eqn:nonconst_a}). In the case of the constant acceleration wave, the initial fit to the measurements and the bootstrapped distributions of initial velocity and acceleration values are shown in Fig.~\ref{cad_boot_weight}. Bootstrapping in this manner allows the determination of confidence intervals on the fit parameters. This is taken from the 100$\alpha$th and 100$\left(1-\alpha\right)$th percentiles of the distribution (giving a 95\% confidence interval when $\alpha=0.025$). Since unknown sources of error can exist that affect the measurement scatter, the only uncertainty that can be confidently attributed to the measurements is that which is due to the resolution limit of the data and similar quantifiable sources of uncertainty. Therefore, in an effort to avoid assigning incorrect uncertainties that will undermine the interpretation of any subsequent model-fit, the scatter of the data instead may be investigated via resampling methods.

For the simulated measurements in Fig.~\ref{cad_boot_weight}, the data points were assigned errorbars of $\pm1.1$\,Mm to represent the lower limit of quantifiable uncertainty, based on the 1.5\,arcsec resolution of AIA. If, as in this simulated coronal wave case, it may be assumed that the model should give an appropriate fit to the data, and the measurement uncertainties are known to be too small (or alternatively too large), then the output uncertainties on the fit parameters can be rescaled accordingly, as per the following equation,
\begin{equation}
\sigma'^2 = \sigma^2 \frac{\chi^2}{\nu} \ ,
\end{equation}
where $\chi^2$ is the standard measure of goodness-of-fit and $\nu$ is the number of degrees of freedom in the fit \citep[see, for example,][]{2003drea.book.....B}. The rescaled uncertainties for this case are quoted in the top panel of Fig.~\ref{cad_boot_weight}. However, since in truth we do not know the exact kinematic form a CME or coronal wave should take nor the true uncertainty due to possible unknown sources of error, we cannot make such assumptions in our treatment of the real-data measurements. Therefore, the power of a bootstrapping technique is clear, by allowing an appropriate confidence interval to be assigned to the fit parameters. For the simulated measurements in Fig.~\ref{cad_boot_weight}, it is seen that the bootstrapped distributions for the initial velocity and acceleration do contain the true model values and have a range similar to the rescaled uncertainties of the single fit (i.e., likening the 95\% confidence intervals to $2\sigma$ uncertainty ranges, which are approximately double the rescaled $1\sigma$ uncertainty ranges of the single fit).

\begin{figure}[t]
\centering
\subfigure{\includegraphics[scale=0.6, trim=0 80 0 30]{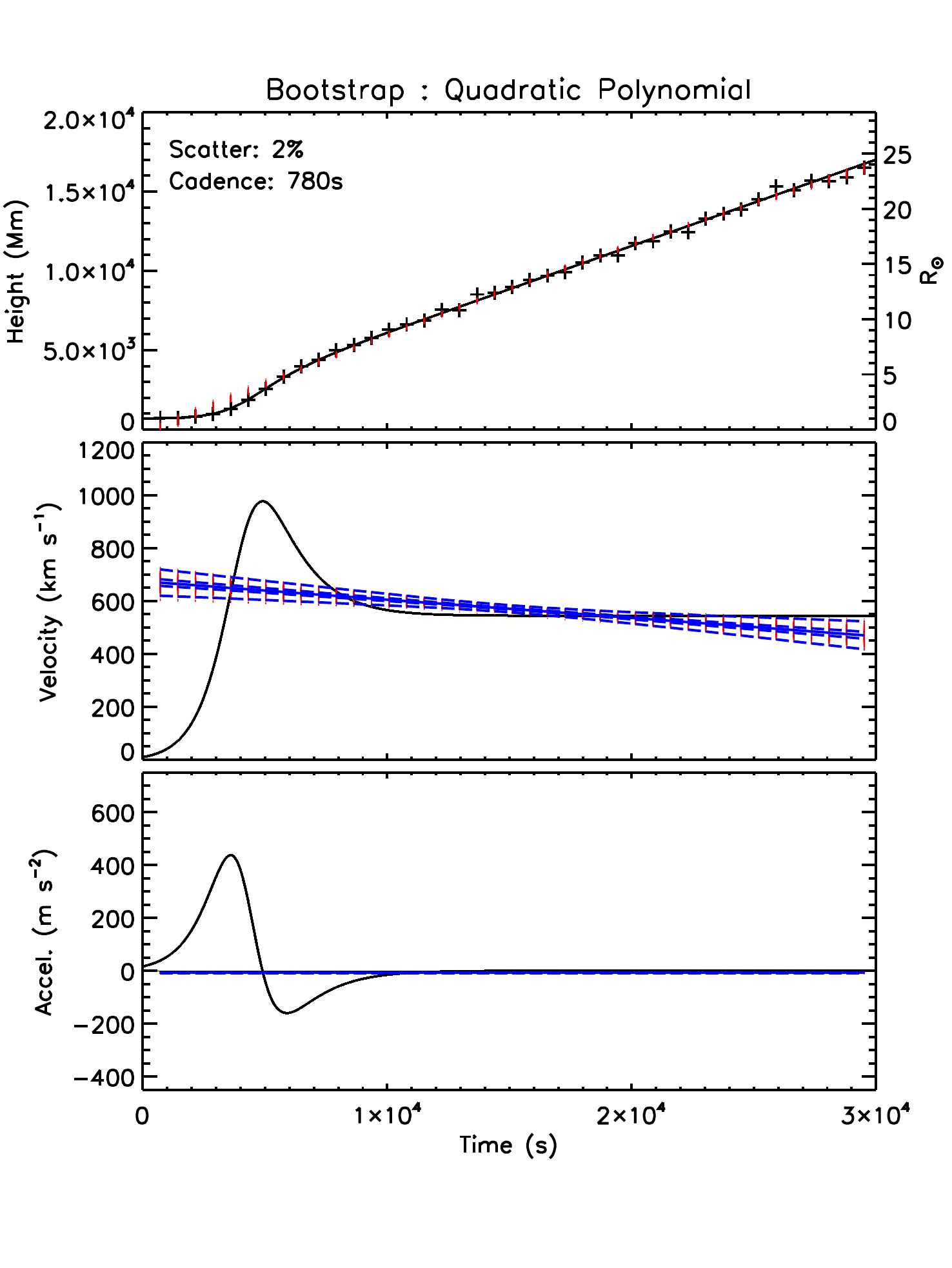}}
\subfigure{\includegraphics[scale=0.6, trim=0 20 0 0]{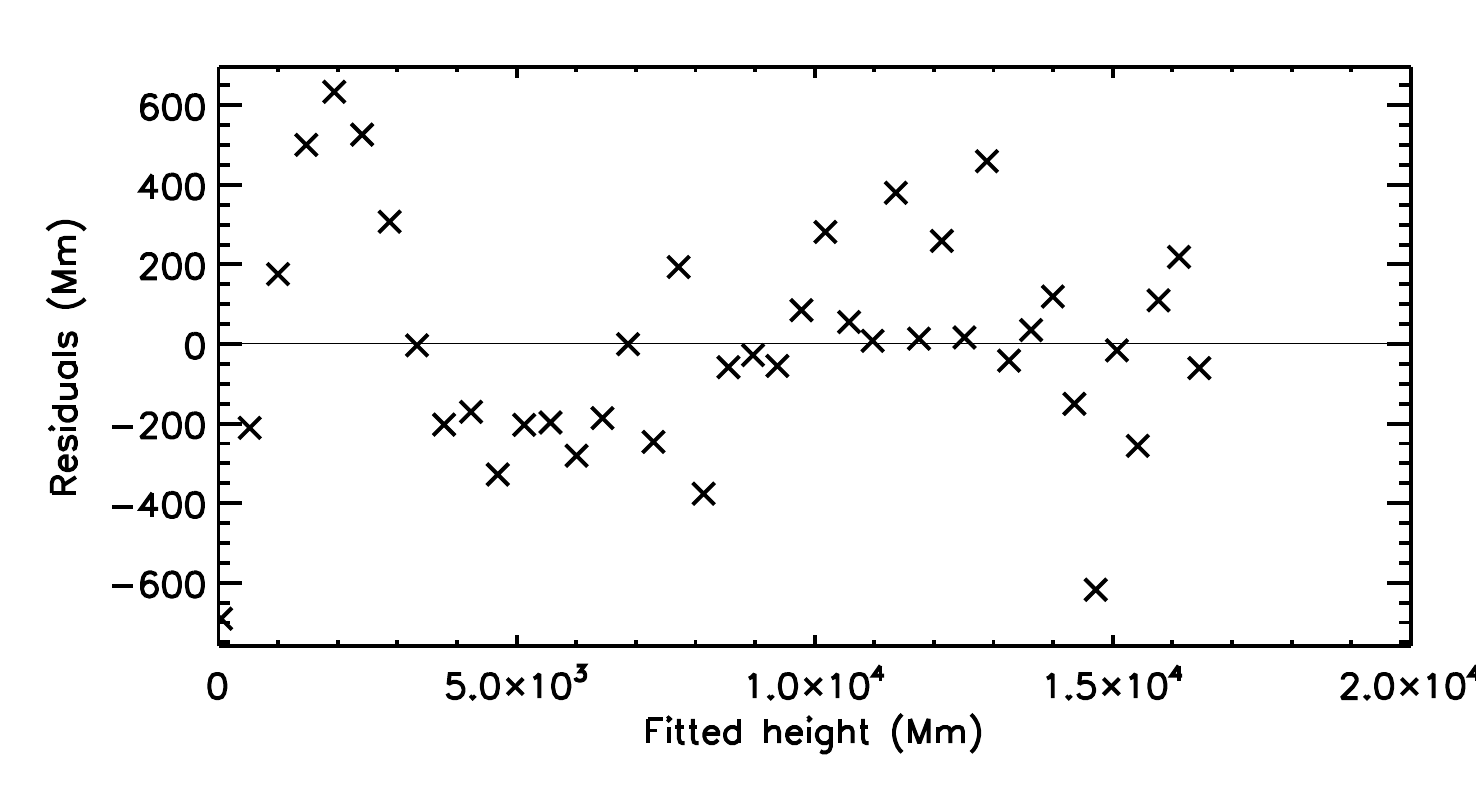}}
\caption{Bootstrapped second order polynomial fit to simulated CME height-time measurements, with 2\% scatter and 780\,s cadence. The panels from top to bottom show the height, velocity, and acceleration plots, and the residuals of the initial fitted height. The red points show the resampled residuals with replacement, and the blue dashed lines are the median, interquartiles range, and upper and lower fences on the bootstrapped fit. The quadratic form tends to smooth out the non-constant acceleration profile, as revealed by the trend in the residuals, indicating that the fit is not appropriate for the measurements.}
\label{fig_quadratic}
\end{figure}

\begin{figure}[t]
\centering
\subfigure{\includegraphics[scale=0.6, trim=0 80 0 30]{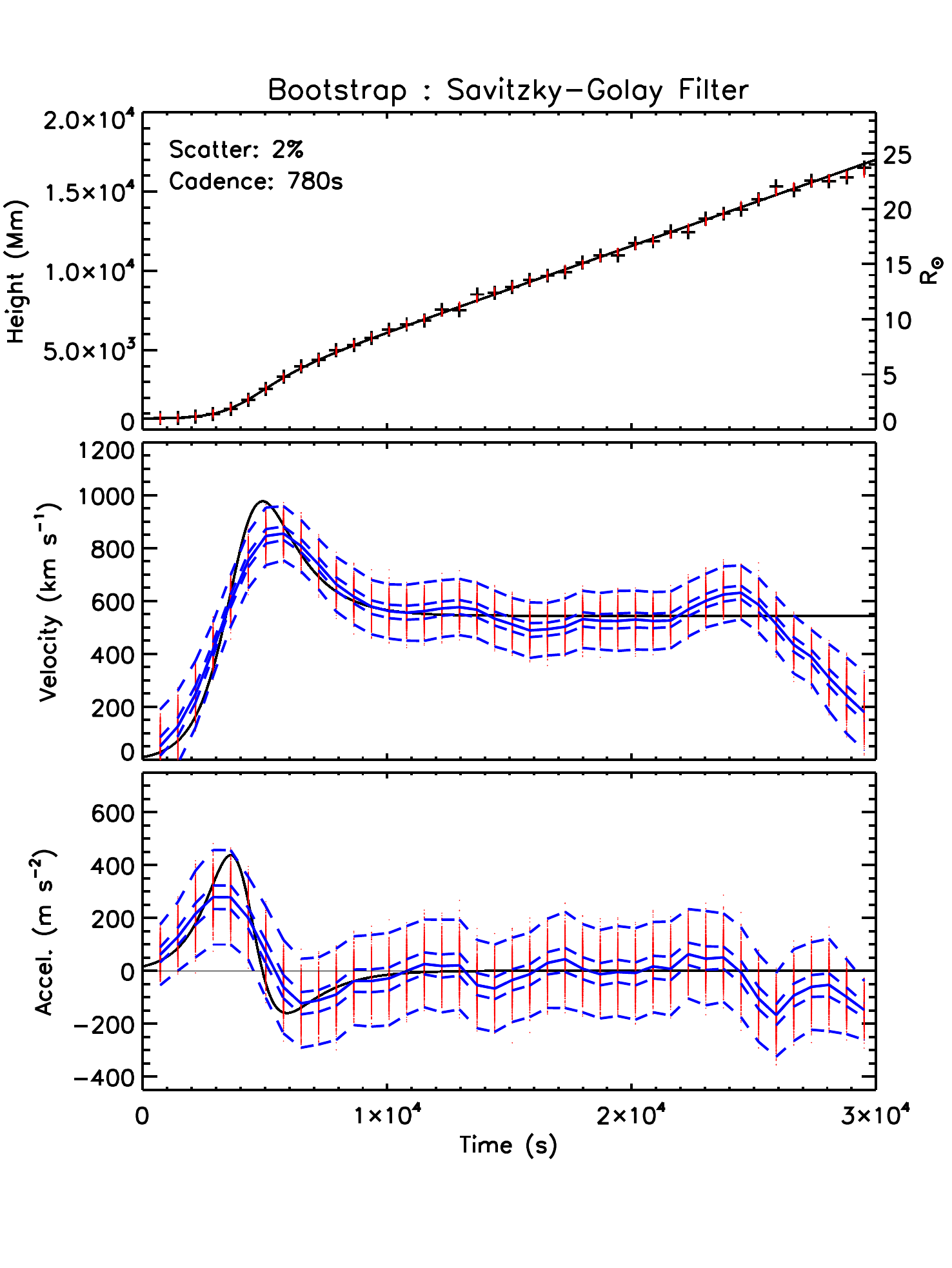}}
\subfigure{\includegraphics[scale=0.6, trim=0 20 0 0]{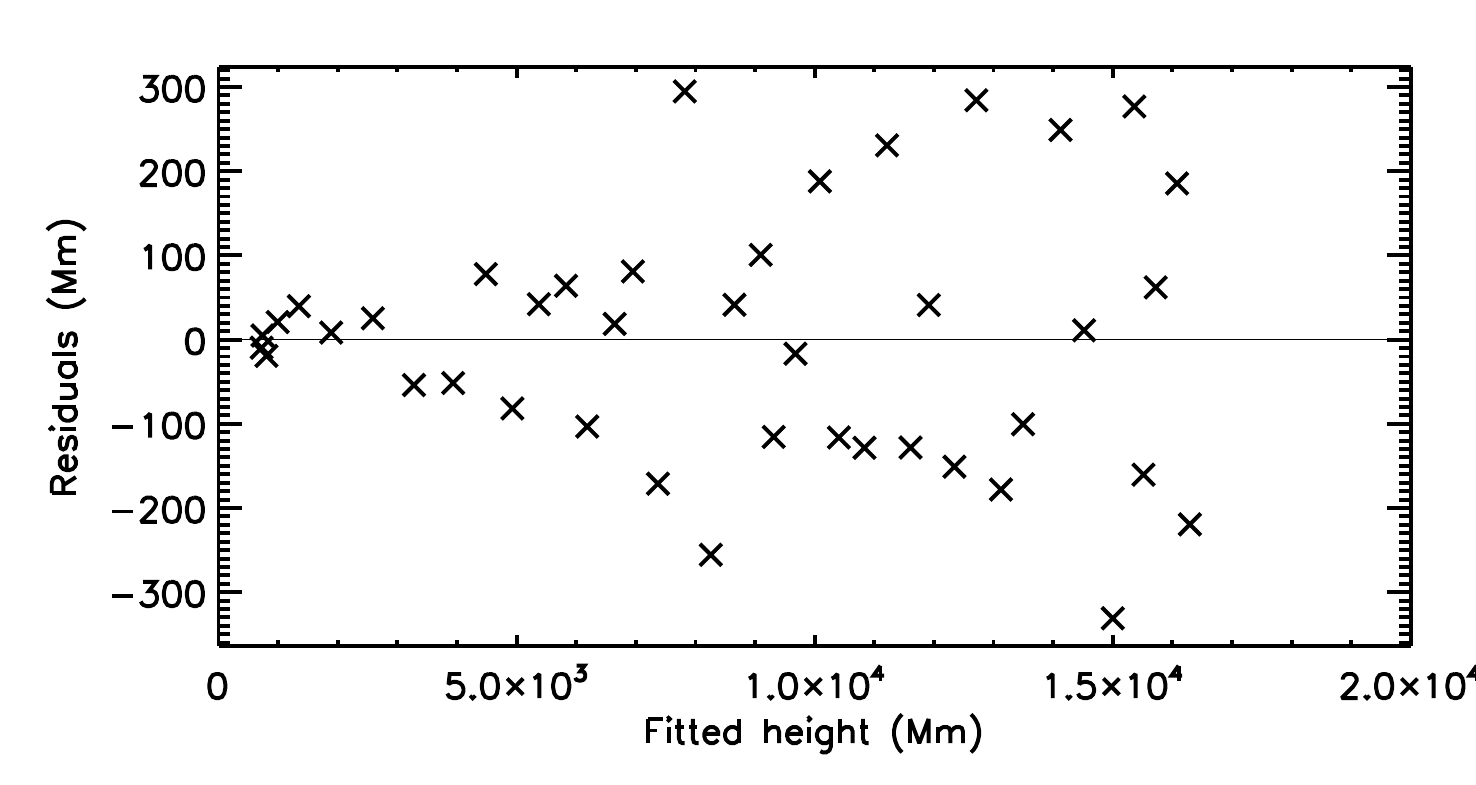}}
\caption{Bootstrapped Savitzky-Golay filter method applied to the simulated CME height-time measurements as in Fig.~\ref{fig_quadratic}. This manner of piecewise fit smooths the measurements by fitting a polynomial to the 3 neighbouring points either side of each data point, and is successful in revealing the non-constant acceleration profile. The randomly scattered residuals also indicate its appropriateness.}
\label{fig_savgol}
\end{figure}

Bootstrapping of a model cannot be applied blindly, as we demonstrate for a simple case of fitting the same constant acceleration form of Equation~\ref{eqn:const_a} to the non-constant acceleration of Equation~\ref{eqn:nonconst_a}. Figure~\ref{fig_quadratic} shows a model non-constant acceleration CME profile sampled at 780\,s cadence with 2\% scatter. The red points show the resulting distributions of points after the residuals are resampled with replacement. Thus a distribution of velocity and acceleration values are derived at each data point, with the corresponding median, interquartile range, and upper and lower fences overlaid in blue\footnote{The interquartile range is the difference between the upper and lower quartiles of a dataset ($IQR=Q_3-Q_1$), determined by the 25th and 75th percentiles, with the second quartile $Q_2$ being the 50th percentile, which is the statistical median of the dataset. Outliers are determined by the upper and lower fences of the dataset: $Lower Fence=Q_1-1.5(IQR)$; $Upper Fence=Q_3+1.5(IQR)$.}. The second-order model is not appropriate to the true non-constant acceleration profile, as revealed by the trend in the residuals of the initial fit (bottom panel of Fig.~\ref{fig_quadratic}). So, for any cases where possible non-constant acceleration profiles are to be revealed, the method for deriving the kinematics must be applied at an appropriate scale such that the residuals scatter randomly. Ideally, a piecewise function should be used to characterise the different phases of motion. This is inspected in great detail by \citet{2008ApJ...674..586S} in an effort to model the early acceleration phase of erupting filaments involved in CMEs. They show that one functional form alone cannot describe the entire phase as well as a number of different functions can. We must rely on some form of numerical derivative for revealing possible trends because we do not know the functional form that CME or coronal wave kinematics will have.

Since the issues with the 3-point Lagrangian have been highlighted in Sect.~\ref{sect:simul1}, we shall opt instead to implement the Savitzky-Golay filter \citep{Savitzky-Golay1964}. This is a form of local polynomial regression of chosen degree of smoothing polynomial, and of chosen order to produce smoothed first order, second order, etc., derivatives of the signal. The number of data points either side of the case point to be included in the filter is also specified. Therefore, in the case of CME and coronal wave distance-time measurements, the Savitzky-Golay filter is a better method for smoothing small-scale scatter while still revealing the true kinematic profiles. This is illustrated in Fig.~\ref{fig_savgol} for the same simulated non-constant acceleration CME profile used in Fig.~\ref{fig_quadratic}, with the residuals resampled with replacement as per the bootstrapping technique described above. For this case, the neighbouring 3 points to the left and right of each data point were considered, and the filter applied with a chosen order of 2. Since it is a form of ``moving window averaging", slight biases may be introduced at local maxima and minima where the function value can be reduced, but its implementation still proves more robust than the standard 3-point Lagrangian. Note that the residuals in the bottom panel of Fig.~\ref{fig_savgol} can be seen to scatter somewhat randomly, as desired. Also, for this realisation of the model it is important to note how the scatter at the endpoints gives the impression of a decreasing velocity which the Savitzky-Golay filter faithfully reproduces even though we know it is not how the model is behaving. This is similar to how the intensity of a CME or coronal wave often lies too close to the background intensity at such distances, being lost to the noise and causing the measured profile to drop off. This alludes to considerations that must be made when dealing with automated systems of kinematic determination, whereby the algorithmic limitations can introduce systematic biases not accounted for in the derived kinematics and associated uncertainties. This is discussed further in the next section.

\section{Case studies}
\label{sect:case_studies}

In this section, some examples of real data are presented as case studies for deriving the CME and coronal wave kinematics in light of the discussion of the previous sections. 

\subsection{CME kinematics}
\label{subsect:corimp}

First, we shall revisit a CME studied by \citet{2009A&A...495..325B} that was observed by \emph{SOHO}/LASCO on 2000 January 2. In that study, the CME front edge was detected via multiscale methods and characterised with an ellipse-fit that was used to track changes to the CME front over time. The apex of the fit (furthest distance from Sun-centre) was measured in each frame and a height-time profile produced. This allowed an investigation into the kinematics of the event, derived using the 3-point Lagrangian method and associated error propagation formulation. However, as has been outlined in the preceding sections, this formulation is somewhat redundant for such a small sample size, and so a new method is applied to test the validity of the analysis.

\begin{figure}[!t]
\centering
\includegraphics[scale=0.6, trim=0 40 0 30]{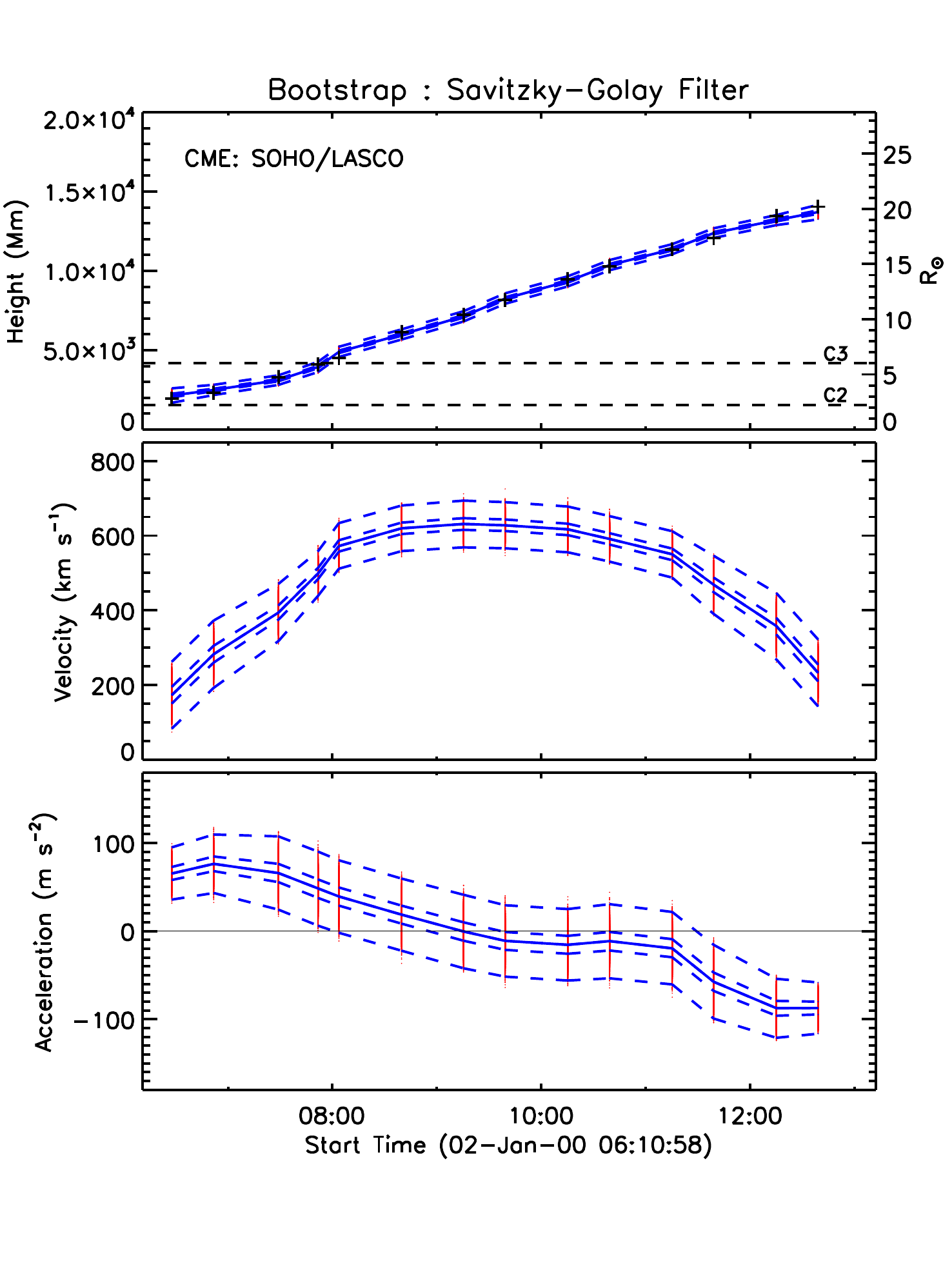}
\caption{Bootstrapped Savitzky-Golay filter method applied to a CME event observed by \emph{SOHO}/LASCO on 2000 January 2, revisited from \citet{2009A&A...495..325B}. The top plot shows the height-time measurements (\emph{plus symbols}), the resampled residuals (\emph{red points}), and the median (\emph{solid line}), interquartile range (\emph{inner dashed lines}), and upper and lower fences (\emph{outer dashed lines}). The middle and bottom plots show the corresponding velocity and acceleration profiles.}
\label{fig_savgol_CME}
\end{figure}

The method chosen, and deemed most appropriate from our investigations in this paper, is that of the Savitzky-Golay filter and bootstrapping technique. The filter width is set at 7 neighbouring data points (i.e., 3 points either side of the point in question). The top plot of Fig.~\ref{fig_savgol_CME} shows the height-time plot of the CME (plus symbols), the resampled residuals after applying the Savitzky-Golay filter (in red), and corresponding median, interquartile range, and upper and lower fences on the bootstrapped data (blue lines). The middle and bottom plots show the derived velocity and acceleration profiles. The variation of the velocity profile is consistent with that of \citet{2009A&A...495..325B}, however the improvement offered by the approach here is the ability to investigate the acceleration profile, which was not possible to perform in \citet{2009A&A...495..325B} due to the unrealistically large errorbars resulting from the numerical differencing technique.

An important point to note is that a priori knowledge about such events and the manner in which they are tracked must be called upon when interpreting the derived kinematic profiles. Edge effects in the derivation of the kinematics can be problematic at both ends of a time-series, though often less so at the start for these types of observations when the accuracy and precision are relatively high before the structure becomes too faint and/or disjoint. A caveat on any method for determining the kinematics is that the coverage of the observations may not be sufficient, either spatially or temporally, to confidently reveal a true trend in the profiles rather than succumbing to edge effects. For CMEs, the overlap between coronagraphs can cause discrepancies in the derived kinematics, and the outer edge of the field-of-view is inherently less reliable due to increased noise as the CME intensity falls-off. For this example, there appears to be a decrease in velocity (and negative acceleration) at the outer edge of C3, but by inspecting the resampled height-time measurements this may be deemed an artefact of the smoothing by the Savitzky-Golay filter when dealing with the scatter on the endpoints -- an ongoing issue for small data sets. This strongly indicates the need for improved CME observations through increased imager cadences and greater field-of-view coverage, but otherwise a spread of measurements across the full span of a CME should be considered in current observations.

\begin{figure}[!t]
\centering
\subfigure{\includegraphics[scale=0.195, trim=0 90 150 0]{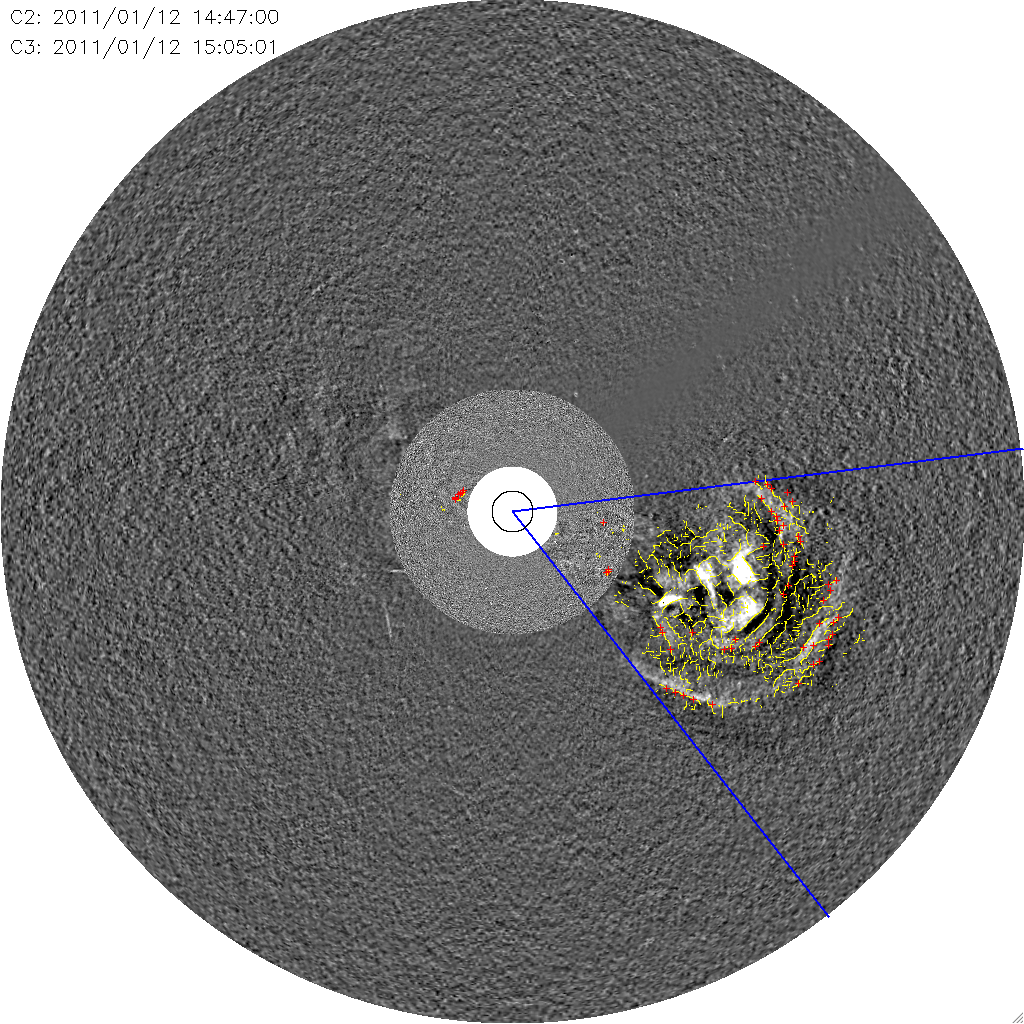}}
\subfigure{\includegraphics[scale=0.465, trim=15 20 0 60]{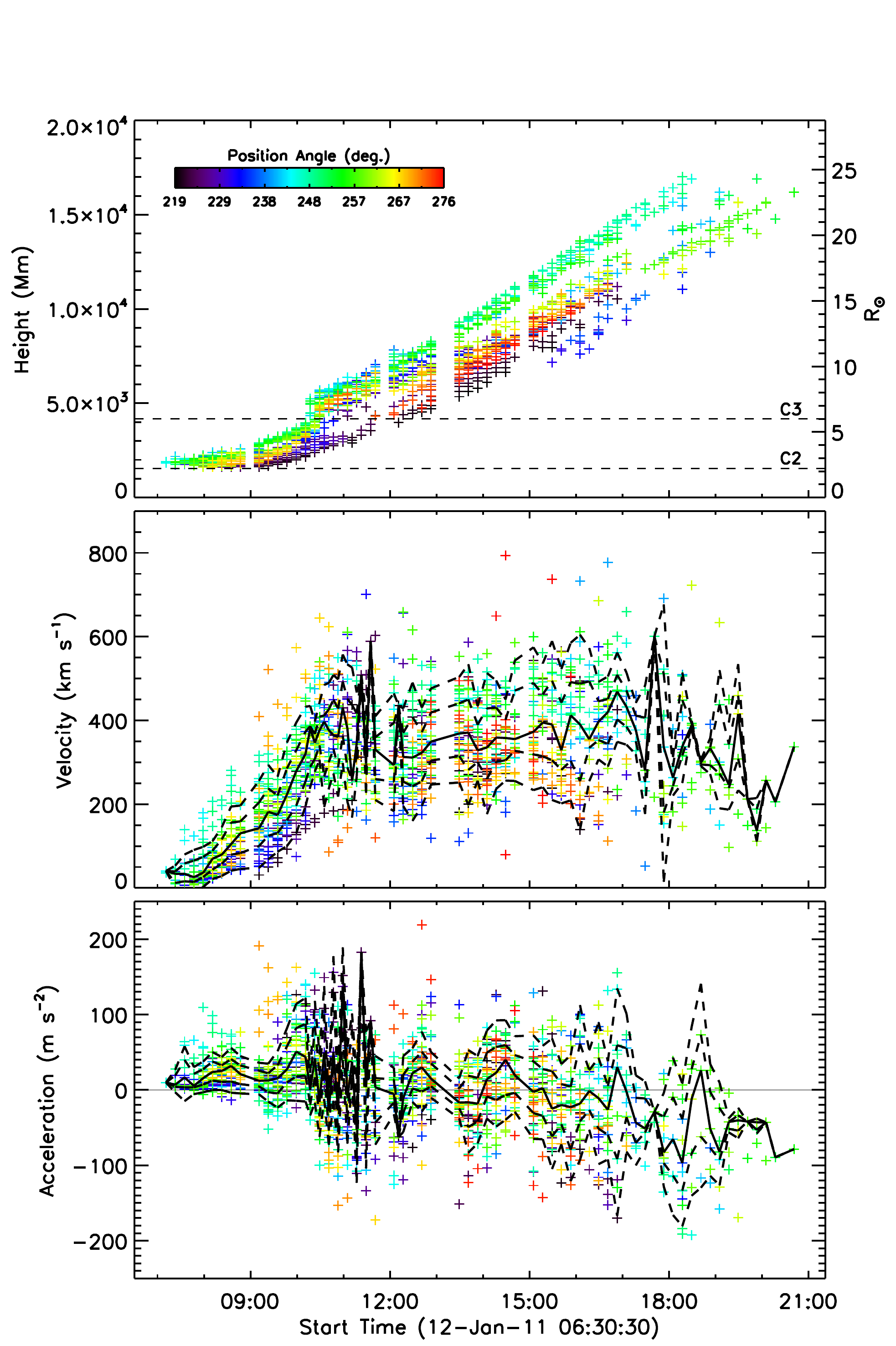}}
\caption{Savitzky-Golay filter applied to the automated CORIMP CME detection and tracking of an event observed by \emph{SOHO}/LASCO on 2011 January 12 (\emph{top image}). The detected CME structure is highlighted in yellow, and the outermost height measurements indicated in red. The top plot of the lower panels shows the height-time measurements across the angular range of the CME (indicated by the colourbar). The middle and bottom plots show the derived velocity and acceleration profiles, with the median (\emph{solid line}), interquartile range (\emph{inner dashed lines}) and upper and lower fences (\emph{outer dashed lines}) over-plotted.}
\label{fig_savgol_CME_CORIMP}
\end{figure}

To this end, we next investigate a CME that occurred on 2011 January 2 as a case-study from the newly developed CORIMP CME catalogue \citep{2012ApJ...752..144M, 2012ApJ...752..145B}. Height-time profiles are measured at 1\,degree intervals across the angular span of the event as it moves through the corona in time (see top plot of Fig.~\ref{fig_savgol_CME_CORIMP}). The Savitzky-Golay filter is then applied to each of these slices along the CME path, and the corresponding first and second order derivatives of the fit are determined. This reveals the spread of velocity and acceleration points as plotted in Fig.~\ref{fig_savgol_CME_CORIMP}, with the median, interquartile range, and upper and lower fences over-plotted in order to characterise the significance of the kinematic range. This allows the underlying trend of the data to be inspected, revealing an initial acceleration of approximately 20\,m\,s$^{-2}$ up to a constant velocity of approximately 400\,km\,s$^{-1}$. Large scatter can inevitably occur at the crossover of the two fields-of-view and at the outer edge of C3.

Since a spread of measurements across the CME can be determined in these cases, the sample size is therefore larger and the spread can be treated as the range of variation in the kinematics without necessarily requiring bootstrap methods to be employed. The implication here is that any variation in CME speeds that may result from the expansion of the ejecta across the plane-of-sky should provide a distribution of kinematics that is centred on the motion of the bulk of the CME (as opposed to the relatively slower moving flanks of the CME). In essence, this provides a solution space of CME kinematics that is based purely on the distribution of CME height-time measurements. For the automated catalogue, avoiding the need for bootstrapping in this manner also saves on computing time. However, if a user of the catalogue has a specific model that they wish to test against the output, then a bootstrapping procedure and residual analysis would be of use, as demonstrated in Sect.~\ref{sect:bootstrapping}. Any position angle may be chosen for such analyses, such as the central position angle that may correspond to the maximum speed of the CME. For example, the position angles around 250$^{\circ}$ in Fig.~\ref{fig_savgol_CME_CORIMP} indicate speeds up to approximately 500\,km\,s$^{-1}$, so for these specific height-time measurements a model may be fit and a bootstrapping technique used to provide a confidence interval for testing the goodness-of-fit. Thus, the output produced from this treatment of the kinematics greatly improves the ability to study CME dynamics.

\subsection{Coronal wave kinematics}
\label{subsect:corpita}

\begin{figure}
\centering
\includegraphics[scale=0.6, trim=0 40 0 30]{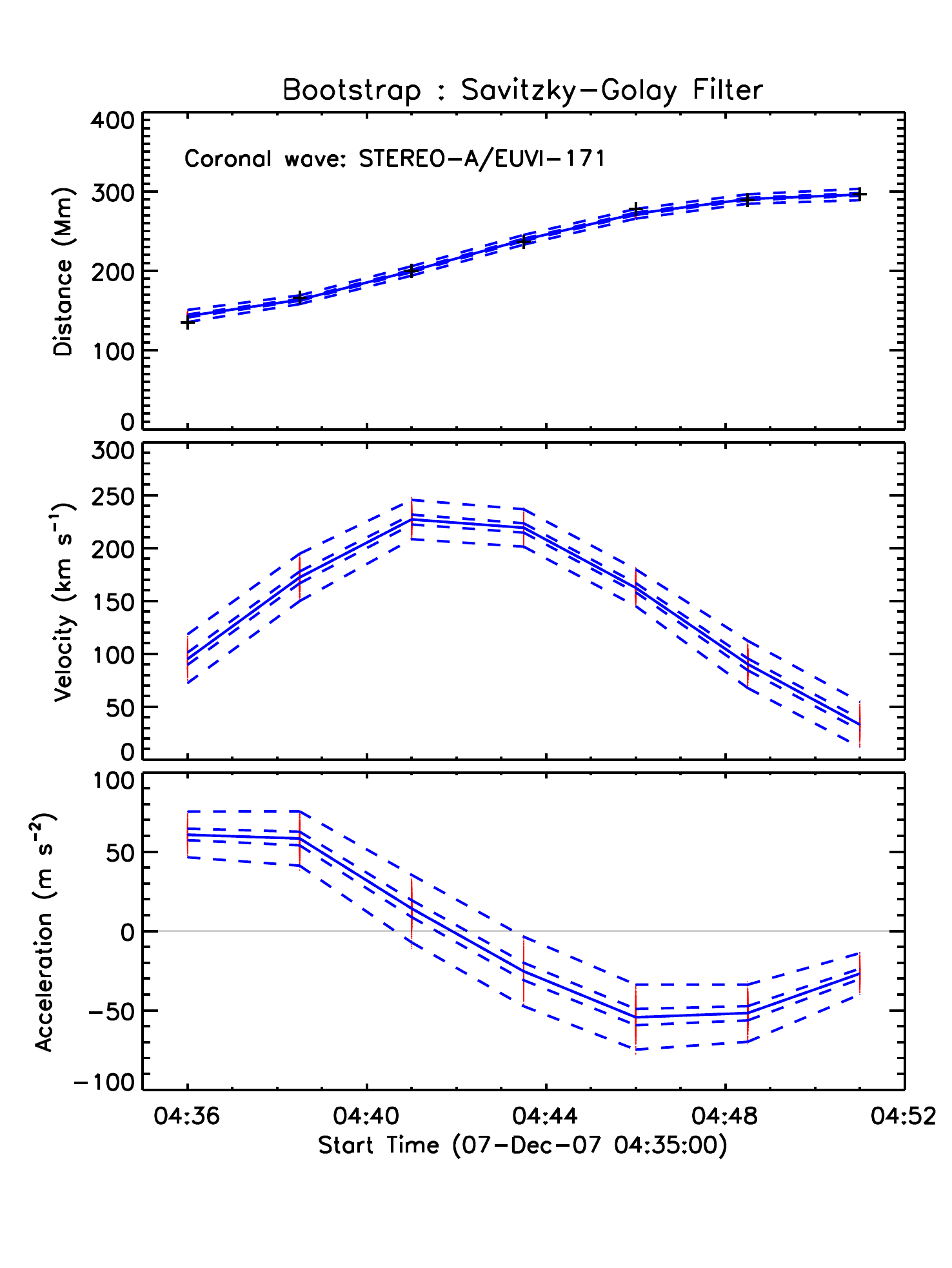}
\caption{Bootstrapped Savitzky-Golay filter method applied to a coronal wave event observed by \emph{STEREO-Ahead}/EUVI 171\AA\ on 2007 December 7, revisited from \citet{2011A&A...531A..42L}. The top plot shows the distance-time measurements (\emph{plus symbols}), the resampled residuals (\emph{red points}), and the median (\emph{solid line}), interquartile range (\emph{inner dashed lines}), and upper and lower fences (\emph{outer dashed lines}). The middle and bottom plots show the corresponding velocity and acceleration profiles.}
\label{fig_savgol_wave}
\end{figure}

\begin{figure}[!t]
\centering
\subfigure{\includegraphics[scale=0.56, trim=10 45 0 75]{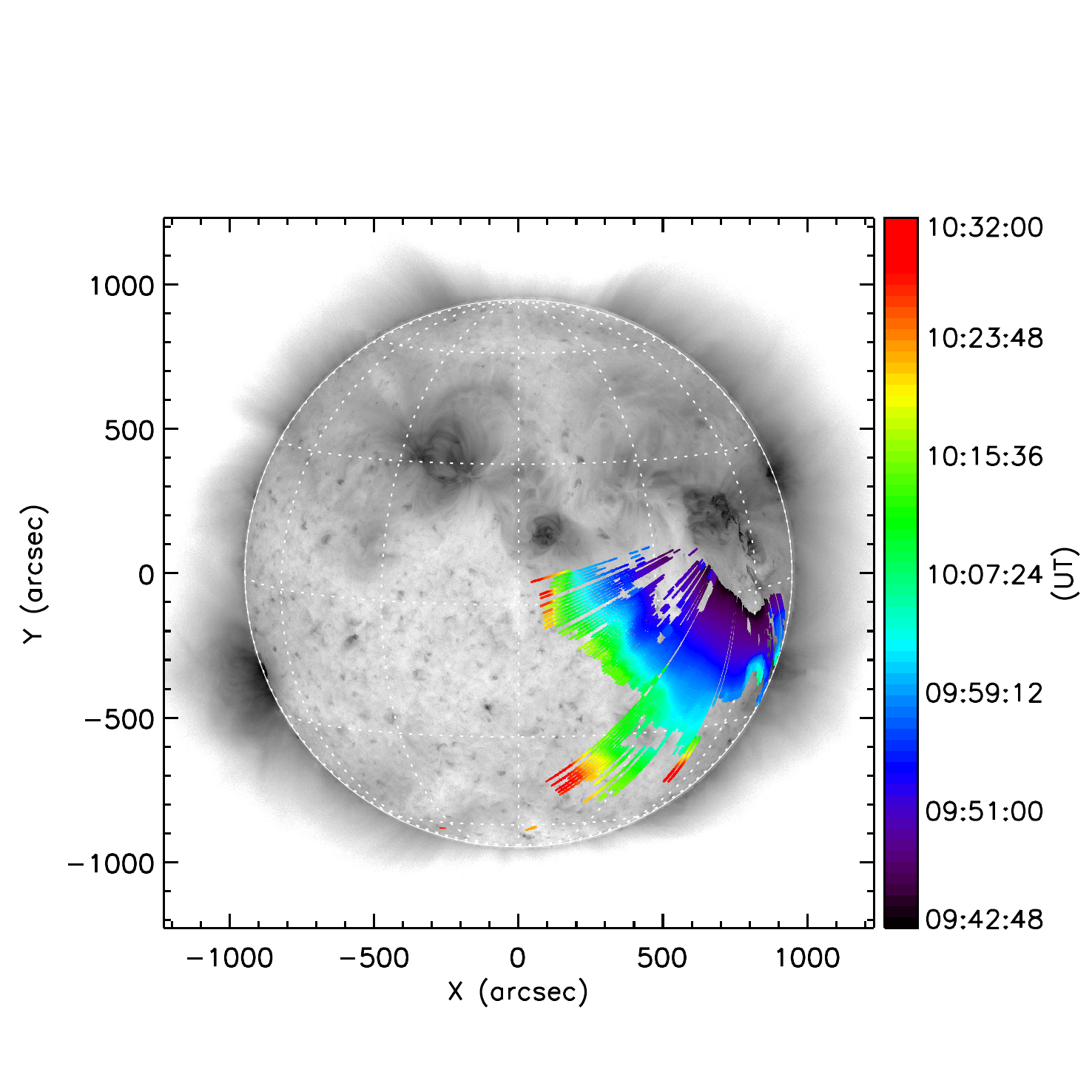}}
\subfigure{\includegraphics[scale=0.46, trim=10 15 0 70]{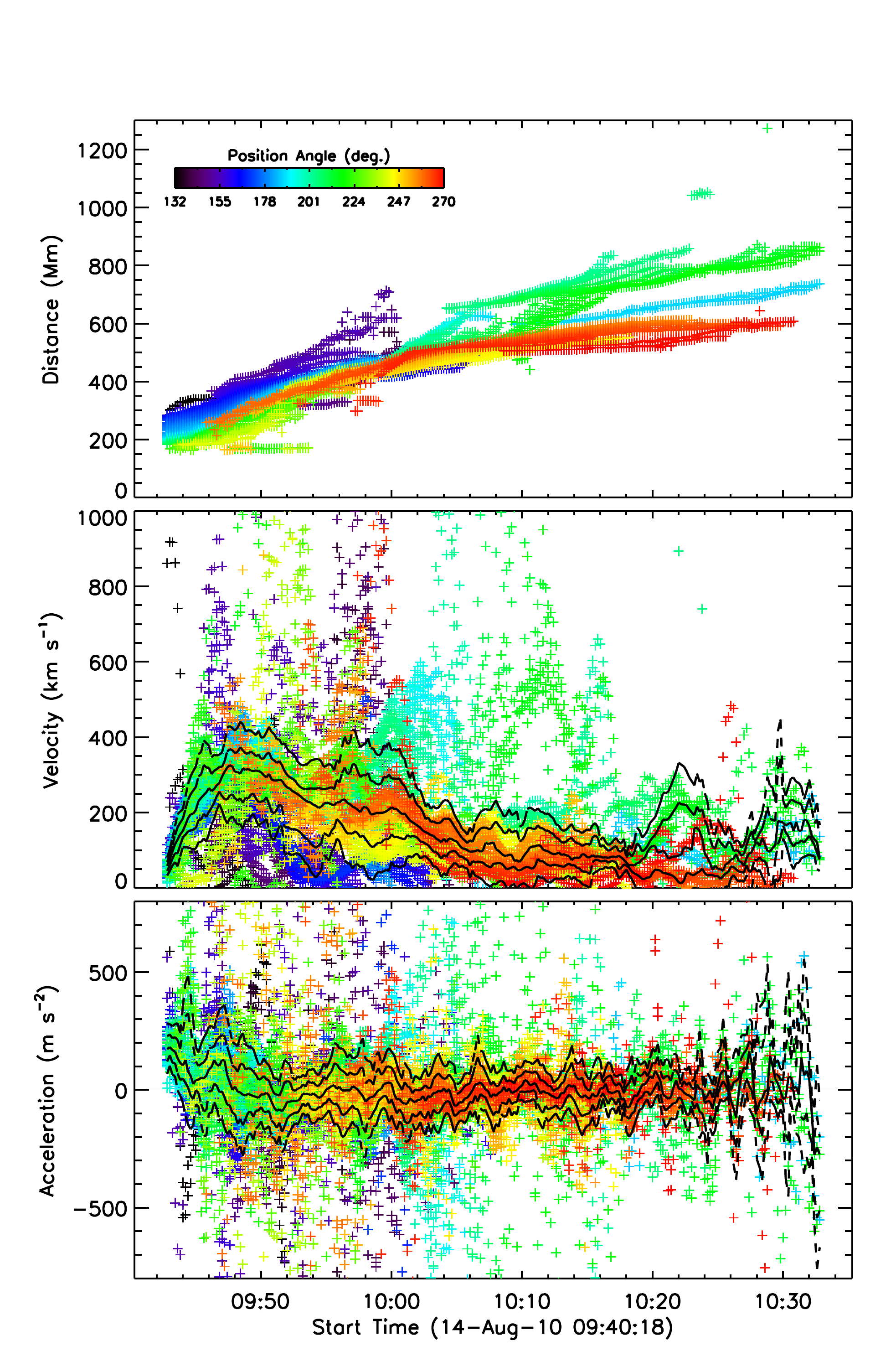}}
\caption{Savitzky-Golay filter applied to the automated CorPITA detection and tracking of a coronal wave observed by \emph{SDO}/AIA on 2010 August 14. The top image shows a percentage base differenced frame during the event, with an overlay of the detected wave motion in time (indicated by the top right colourbar). The top plot of the lower panels shows the distance-time measurements across the angular range of the coronal wave (indicated by the inset colourbar). The middle and bottom plots show the derived velocity and acceleration profiles, with the median (\emph{solid line}), interquartile range (\emph{inner dashed lines}) and upper and lower fences (\emph{outer dashed lines}) over-plotted.}
\label{fig_savgol_wave_CorPITA}
\end{figure}

The treatment of coronal wave kinematics is very similar to that of CMEs, in that a small sample size of distance measurements is obtained, being prone to the effects of scatter from observational and algorithmic biases. We first revisit a coronal wave event studied by \citet{2011A&A...531A..42L}, that was observed by the Solar Terrestrial Relations Observatory \citep[\emph{STEREO};][]{2008SSRv..136....5K} Extreme-ultraviolet Imager (EUVI 171\AA\ ) on 2007 December 7. In that study, the coronal wave was detected via an intensity thresholding technique on percentage base-difference images. Distance-time measurements were produced and studied with a quadratic model of the form of Equation~\ref{eqn:const_a}. Here, we use the Savitzky-Golay filter and bootstrapping technique, as in the CME case study of Fig.~\ref{fig_savgol_CME}. The filter width was reduced to operate on 5 neighbouring data points (i.e., 2 either side) since the sample size itself is so small, consisting of only 7 measurements. The top panel of Fig.~\ref{fig_savgol_wave} shows the distance-time plot of the coronal wave (plus symbols), the resampled residuals after applying the Savitzky-Golay filter (in red), and corresponding median, interquartile range, and upper and lower fences on the bootstrapped data (blue lines). The middle and bottom panels show the derived velocity and acceleration profiles. The kinematic trend strongly implies an initial acceleration of approximately 60\,m\,s$^{-2}$ followed by deceleration of $-50$\,m\,s$^{-2}$ as it then approaches zero velocity. It is interesting that the velocity profile of Fig.~\ref{fig_savgol_wave} differs dramatically from that expected of a constant-acceleration kinematic profile, as was fit in \citet{2011A&A...531A..42L}. Care must still be taken when interpreting this trend from such a small data set, and such trends would be better justified with increased imager cadences, as is now possible with \emph{SDO}/AIA.

The benefit of a higher cadence was demonstrated in Fig.~\ref{cad_hist_weight} for the 12 second cadence of AIA and, since the resolution of the images is also significantly higher than previous missions, the determination of the true kinematics should be greatly improved as a result (c.f. Fig.~\ref{noise_test_image}). A coronal wave event observed by AIA, and tracked via the automated CorPITA algorithm, is shown in Fig.~\ref{fig_savgol_wave_CorPITA}. The top image and upper plot show the distance-time measurements, and middle and bottom plots show the derived velocities and accelerations, respectively. The Savitzky-Golay filter was again deemed optimal for determining the kinematics of the event, at a filter width of 11 neighbouring data points (i.e., 5 either side). Similar to the previous CME case, this provides a spread of kinematics that is centred on the motion of the bulk of the event, with the median, interquartile ranges and upper/lower fences over-plotted. The trend indicates a possible initial acceleration of 100--150\,m\,s$^{-2}$, attaining speeds of approximately 300\,km\,s$^{-1}$ before slowing down. The general limitation to be highlighted here is that the interpretation of the kinematics along any single position angle can result in different profiles that would imply different wave dynamics. A spread of measurements must be obtained in order to best characterise the eruption as a whole, noting that the scatter provides an insight into the variability of the results. Any position angle may be chosen for specific investigation of whether the detection technique has introduced erroneous trends, or is in fact revealing the different physics underlying the propagation of any given event.

\section{Conclusions}
\label{sect:conclusions}

We have demonstrated the unreliability of some common approaches to studying the kinematics of large scale dynamic features in the solar corona, such as CMEs and coronal waves. The dynamics of these phenomena are of great consequence to understanding the physics governing their initiation and propagation, and so the drawbacks of traditional numerical techniques for deriving their velocities and accelerations have been highlighted and efforts to overcome them discussed. Of particular note is the counter-intuitive behaviour of the error propagation in the standard 3-point Lagrangian interpolation, due to its inverse dependence on the observing cadence. It is also shown how strongly affected it can be by noise levels similar to those found in the data sets of these events. Therefore, we outline a two-fold approach.

Firstly, we suggest the use of bootstrapping techniques when dealing with the small sample sizes of these events. We demonstrate both their ease of implementation and usefulness in estimating the confidence intervals on fit parameters. It is an effective method for overcoming the limitation of having only a single set of observations of a phenomenon, which is the case for the dynamic events studied here. As a class of resampling methods, bootstrapping allows the construction of a number of resamples of the observed data from which a distribution of fitting parameters may be generated. Thus, the most likely parameter values and associated uncertainties may be determined. Uncertainty estimates calculated via a defined and appropriate procedure are crucial for determining the appropriateness of any chosen theoretical model seeking to characterise the physics of their motion. A robust estimate of the goodness-of-fit can be obtained when used in conjunction with residual analysis, as demonstrated in Figs.~\ref{fig_quadratic} and \ref{fig_savgol}. This is because bootstrapping itself cannot be applied blindly, which leads us to the next point.

Secondly, we suggest that the events may be treated with a priori knowledge since there is a physical basis for their behaviour that can be confidently assumed to a certain extent. For example, the plasma is known to start from rest, and thus undergo an early acceleration to reach its maximum speed. What remain to be tested are the exact form of that initial acceleration phase, whether or not unexpected trends or changes occur in its motion, and whether additional effects come into play (such as from drag and/or interactions with other features of the surrounding environs). The interpretation of the measurements must also allow for unknown biases that can occur, either due to observational uncertainties (e.g., low intensity profiles embedded in background noise) or algorithmic uncertainties (e.g., erroneous scatter derived at the boundary between instrument fields-of-view). Thus, while the uncertainty of a measurement might be mathematically sound, the measurement itself could be offset in a manner the user must aim to interpret appropriately in the derived kinematics, as demonstrated in Fig.~\ref{fig_savgol_CME}.

Many of the issues of traditional numerical differencing techniques and standard error propagation can be overcome by the use of more appropriate methods. Thus, we may improve the determination of CME and coronal wave kinematics, and indeed this extends to any similar scenario of a dynamic event. We note that the significance attributed to the kinematic profiles must still be interpreted with care, relying on a priori knowledge so that none of these techniques be blindly applied or interpreted. This is essentially the idea behind Bayesian inference, wherein the probability of a hypothesis is updated as additional evidence is learned. Techniques derived from Bayesian statistics may prove useful for studying the kinematics of solar dynamic phenomena as observations and catalogues of their properties are built-up over time. For example, \citet{2010SoPh..262..495B} employ a Kalman filter in their development of a dynamic model for three-dimensional tomography of the solar corona. This particular type of algorithm operates recursively (over two steps of prediction and update) to build an estimate of the state of a system and associated uncertainties, having prior knowledge of the state of the system. Such approaches may hold promise if they can be adapted for the treatment of solar event kinematics.

\begin{acknowledgements}
This work is supported by SHINE grant 0962716 and NASA grant NNX08AJ07G to the Institute for Astronomy. The \emph{SOHO}/LASCO data used here are produced by a consortium of the Naval Research Laboratory (USA), Max-Planck-Institut fuer Aeronomie (Germany), Laboratoire d'Astronomie (France), and the University of Birmingham (UK). \emph{SOHO} is a project of international cooperation between ESA and NASA. \emph{SDO} data supplied courtesy of the NASA/\emph{SDO} consortia. The \emph{STEREO}/SECCHI project is an international consortium of the Naval Research Laboratory (USA), Lockheed Martin Solar and Astrophysics Lab (USA), NASA Goddard Space Flight Center (USA), Rutherford Appleton Laboratory (UK), University of Birmingham (UK), Max-Planck-Institut f\"{u}r Sonnen-systemforschung (Germany), Centre Spatial de Liege (Belgium), Institut d'Optique Th\'{e}orique et Appliqu\'{e}e (France), and Institut d'Astrophysique Spatiale (France). The authors would like to acknowledge the help and guidance from the \emph{SDO} Feature Finding Team. DSB was supported by a Marie Curie Intra-European Fellowship under the European Community's Seventh Framework Programme (FP7) and the ESA Prodex program. DL would like to thank the Institute for Astronomy at the University of Hawaii for their generous hospitality during his visit. The authors thank the anonymous referee for their valuable comments.
\end{acknowledgements}

\bibliographystyle{apj.bst}
\bibliography{references.bib}

\end{document}